%% file: draft.tex
\documentclass[letterpaper,twocolumn,10pt]{article}

\usepackage{usenix-2020-09}

\usepackage{hyperref}
\usepackage{graphicx}
\usepackage{wrapfig}
\usepackage{sidecap}
\usepackage{color}
\usepackage{xcolor}
\usepackage{pifont}
\usepackage{upquote}
\usepackage{subcaption}
\usepackage{proof}
\usepackage{combelow}
\usepackage{blindtext}
\usepackage[most]{tcolorbox} 
\usepackage[export]{adjustbox}
\usepackage{xspace}
\usepackage{booktabs}
\usepackage[frozencache]{minted2}
\usepackage{caption}
\usepackage{enumitem}
\usepackage{cleveref}
\usepackage{longtable}
\usepackage{oplotsymbl}
\definecolor{block-gray}{gray}{0.95}
\definecolor{darkgreen}{rgb}{0.0, 0.5, 0.0}
\usepackage{xurl}
\hypersetup{breaklinks=true,
            colorlinks=true,
            citecolor=blue,
            urlcolor=blue,
            pdfborder={0 0 0}}
\newtcolorbox{prompt-box}[1]{
  fonttitle=\bfseries,
  fontupper=\small,
  breakable,
  enhanced,
  title=#1
}

\newtcolorbox{prompt-box-small}{
  breakable,
  enhanced,
  top=0.1pt,
  bottom=0.1pt,
}

\crefformat{section}{§#2#1#3}
\crefrangeformat{section}{\S#3#1#4--\S#5#2#6}

\ifdefined\vv 
\renewcommand{\vv}{~\vert~}
\else
\newcommand{\vv}{~\vert~}
\fi

\crefformat{figure}{Fig.~#2#1#3}
\crefformat{table}{Tab.~#2#1#3}
\crefformat{section}{\S#2#1#3}
\crefmultiformat{section}{\S#2#1#3}{--\S#2#1#3}{, \S#2#1#3}{ and \S#2#1#3}
\crefrangeformat{section}{\S#3#1#4--#5#2#6}

\def\node{Node.js\xspace}
\newcommand{\sysp}{Lexo\xspace}
\newcommand{\sys}{{\scshape{}\sysp}\xspace}

\def\myomit#1{}
\def\eg{{\em e.g.}, }

\def\etc{{\em etc.}\xspace}

\newcommand{\heading}[1]{\vspace{2pt}\noindent\textbf{#1}:\enspace}

\newcommand{\ttt}[1]{\texttt{#1}}

\newcommand{\tcn}[1]{}

\newcommand{\sx}[1]{(\S\ref{#1})}
\newcommand{\cf}[1]{(\emph{Cf}.\S\ref{#1})}

\newlength\myboxwidth

\newcommand{\gptmodel}{GPT-5 mini\xspace}

\newcounter{gregNOC}
\stepcounter{gregNOC}

\definecolor{ao}{rgb}{0.0, 0.5, 0.0}

\newminted{jscode}{xleftmargin=\parindent, linenos, numbersep=5pt, fontsize=\small}

\newcommand{\totalLibs}{147\xspace}
\newcommand{\sysCorrectLibs}{86\xspace}

\begin{document}

% \keywords{supply-chain attacks, large language models, software regeneration, automated code generation}

% <concept_id>10002978.10003022</concept_id>
% <concept_desc>Security and privacy~Software and application security</concept_desc>
% <concept_id>10011007</concept_id>
% <concept_desc>Software and its engineering</concept_desc>

% \ccsdesc[500]{Security and privacy~Software and application security}
% \ccsdesc[500]{Software and its engineering}

% \author{
%   Anonymous Author(s)\\
%   Paper ID: \#2129
% }

\title{
  Lexo: Eliminating Stealthy Supply-Chain Attacks\\ via LLM-Assisted Program Regeneration
}

% Evangelos Lamprou (Brown University & DTU) <evangelos_lamprou@brown.edu>
% Julian Dai (Brown University) <julian_dai@brown.edu>
% Grigoris Ntousakis (Brown University) <gntousakis@brown.edu>
% Martin Rinard (MIT) <rinard@csail.mit.edu>
% Nikos Vasilakis (Brown University) <nikos@vasilak.is>

\author{
  {\rm Evangelos Lamprou}\\
  Brown University
  \and
  {\rm Julian Dai}\\
  Brown University
  \and
  {\rm Grigoris Ntousakis}\\
  Brown University
  \and
  {\rm Martin C. Rinard}\\
  MIT
  \and
  {\rm Nikos Vasilakis}\\
  Brown University
}

% \author{Evangelos Lamprou}
% \affiliation{%
%   \institution{Brown University}
% \email{vagos@lamprou.xyz}

% \author{Julian Dai}
% \affiliation{%
%   \institution{Brown University}
% \email{julian_dai@brown.edu}

% \author{Grigoris Ntousakis}
% \affiliation{%
%   \institution{Brown University}
% \email{gntousakis@brown.edu}

% \author{Martin C. Rinard}
% \affiliation{%
%   \institution{MIT}
% \email{rinard@csail.mit.edu}

% \author{Nikos Vasilakis}
% \affiliation{%
%   \institution{Brown University}
% \email{nikos@vasilak.is}

% \renewcommand{\shortauthors}{E. Lamprou, J. Dai, G. Ntousakis, M. C. Rinard, N. Vasilakis}

\maketitle

\begin{abstract}
    Software supply-chain attacks are an important and ongoing concern in the open source software ecosystem. 
    These attacks maintain the standard
    functionality that a component implements, but additionally hide malicious
    functionality activated only when the component reaches its target
    environment. \sys addresses such stealthy attacks by
    automatically learning and regenerating vulnerability-free
    versions of potentially malicious components. \sys first generates a
    set of input-output pairs to model a component's full observable behavior, which it then
    uses to synthesize a new version of the original component. The new component
    implements the original functionality but avoids stealthy malicious
    behavior. Throughout this regeneration process, \sys consults several
    distinct instances of Large Language Models (LLMs), uses correctness and
    coverage metrics to shepherd these instances, and
    guardrails their results. An evaluation on 100+
    real-world packages---including high-profile stealthy supply-chain attacks---indicates that
    \sys scales across multiple domains, regenerates code efficiently (<30m on
    average), maintains compatibility, and succeeds in eliminating malicious
    code in several real-world supply-chain-attacks---even in cases when a state-of-the-art LLM 
    fails to eliminate malicious code
    when given the source code of the component 
    and prompted to do so.
\end{abstract}

\section{Introduction}
\label{sec:introduction}

% Thesis: Using LLM-assisted learning and regeneration, humanity can defend against stealthy software supply-chain-attacks without using any kind of detection mechanisms or sandboxing.

% Key Result: 100+ libraries with 100M weekly downloads commutitavely can be regenerated and thus seize being the target of sneky software supply chain attacks.

% Context
% Problem
% The Approach
    % Immediate Objections
% Evaluation overview / broader significance
% 3–4 key contributions 
% Limitations
% Availability

Supply-chain attacks have been an ongoing issue in the software 
industry~\cite{cloudflare2023, sonatype2024, sansec_polyfill_2023, ev:eurosec:2022, leftpad, wired_barium_supply_chain, cybersecuritydive_3cx_attack_2024, npm_attack_ethereum},
resulting in over \$46B damages in recent years~\cite{software_supply_chain_cost}.
These attacks target a victim's supplier, exploiting the fact that the victim software depends on software provided by the supplier.
As modern software comprises thousands of dependencies, % FIXME ~\cite{}, 
  it is challenging for developers to audit all dependencies---and particularly so for dependencies nested deeply in the dependency chain.

\begin{figure}[t]
    \centering
    % https://docs.google.com/drawings/d/1u2bGKPxABdI_8dwrTyhcXVkjKJiYexDdxnVt5RuUSF8/edit?usp=sharing
    \includegraphics[width=\columnwidth]{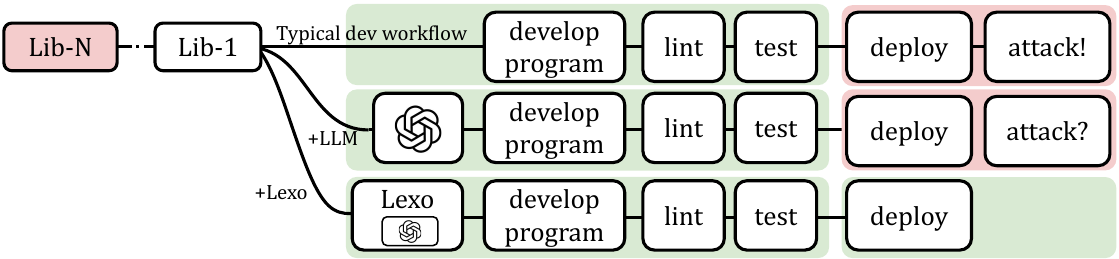}
    \caption{
      \textbf{Applying \sys.}
      Top: A typical workflow today might result in an attack.
      Mid: Applying an LLM to detect or eliminate vulnerabilities might still result in an attack.
      Bottom: \sys regenerates a vulnerability-free version of a component (or reports inability to do so).
    % Application of \sys in comparison to a typical developer wokflow or naively using an LLM to detect or remove vulnerabilities.
    }
    \label{fig:comparison}
    \vspace{-.2in}
\end{figure}

Automated detection of such attacks remains challenging, as malicious components use various evasion techniques.
Specifically, in the case of \emph{stealthy} supply-chain attacks (SSCAs), a malicious dependency may remain dormant until multiple conditions are met---for example, 
  specific values of environment variables,
% the host's operating system, 
  the version of another component,
  the system's date and time, 
  the number of times a component has been invoked, or 
  random chance. These techniques often enable supply-chain attacks to evade detection by standard software validation mechanisms such as testing.

%This vast space of possible conditions can complicate detection. 

% under which a component's malicious behavior could activate 
% makes detection  extremely complex and highly unlikely.
% an exercise in reactive security.

This paper presents \sys, a system that eliminates SSCAs %stealthy supply-chain-attacks
  by combining large language model (LLM)-assisted program inference, transformation, and regeneration.
  \sys does not attempt to directly detect, remove, or interpose on vulnerable or malicious code.
It instead extracts a model of the component's intended behavior (in the form of input/output examples), then uses this model to regenerate a new, secure, vulnerability-free version of the component. A key insight is that a stealthy malicious component still exhibits the advertised client-visible component behavior---which \sys infers, represents as input/output examples, and regenerates. % , discarding the component's malicious behavior.

Because the extracted component model (input/output examples) decouples the
original, potentially malicious or vulnerable, code from the regenerated code,  
\sys is distinct from approaches that present a component directly to the LLM, tasking the LLM to recognize any malicious or vulnerable code to generate a vulnerability-free version of the component~\cf{sec:eval-naive}. 

% using the key insight that a stealthy malicious component still behaves as advertised from a client's perspective,
%While earlier research on vulnerability elimination has demonstrated promising results via active learning and regeneration~\cite{harp:ccs:2021}, 
%The techniques underlying \sys are language- and domain-agnostic---hinting at a potential for regenerating substantial parts of entire ecosystems.
% \sys is the first system that supports scalable library regeneration across domains.

\heading{Deployment scenarios}
\sys produces as output 
  either a vulnerability-free version of a software component it takes as input (Fig.~\ref{fig:system-overview})
  or a message that it cannot operate on its input component---for example, because the original component 
falls outside its regeneration scope.
% produces side-effect as a result of its legitimate computation
% or because \sys cannot guarantee adequate code coverage.

%as an LLM can be tricked, using careful naming patterns).

\sys can be applied at various stages of the software lifecycle.
During development, it can regenerate software dependencies to reduce the risk of developer-targeted attacks.
In continuous integration workflows, it can be applied to release checkpoints to ensure that all dependencies are regenerated before deployment to production environments. 
It can also be applied selectively, regenerating only before release and deployment.
The current version of \sys targets dependencies that sit at the leaf nodes of the dependency graph, as they usually implement well-defined functionality and are ubiquitous across applications within an ecosystem~\cite{ev:eurosec:2022}.

% Additionally, during deployment, for applications that interact with user-facing code, \sys can replace user-created components with safe alternatives at runtime.

\heading{Evaluation overview}
Applied to several real-world supply-chain attacks,
  \sys regenerates new versions of these components
that re-implement the core computation, pass developer tests, support client
applications, and---crucially---eliminate any malicious behavior present in the
original component~(\cref{sec:eval-security}).
\sys correctly regenerates \sysCorrectLibs out of \totalLibs components, confirmed by 100\% of developer-provided tests and by manual inspection, and exits early on the remaining 61, after it detects that the source component falls outside its regeneration capabilities~(\cref{sec:correctness}).
On average, \sys's component regeneration completes in 30m and costs less than \$0.10.

\heading{Paper outline and contributions}
The paper starts with an example of a stealthy supply-chain attack~\sx{sec:example} and the key challenges involved in detection and mitigation.
It then discusses the threat model \sys protects against~\sx{sec:threat-model}, before presenting an overview of the system~\sx{sec:overview} outlining its key contributions:

\begin{itemize}[leftmargin=*,nosep]

    \item \textbf{Program inference:}
        Given a software component, \sys uses the component's source code to generate a series of inputs~\sx{sec:io}.
        It then uses these inputs to create a model that reflects the component's client-observable behavior. \sys then uses this model to regenerate a new, secure version of the component.

    \item \textbf{Language- and domain-agnostic regeneration:}
     \sys introduces an automated, language- and domain-agnostic approach to regenerate
        safe, behaviorally equivalent versions of vulnerable components, eliminating malicious code without human intervention~\sx{sec:synthesis}.

    \item \textbf{Guardrails, optimizations, and refinements:}
        \sys introduces a series of checks for rejecting invalid inputs and incorrectly regenerated components~\sx{sec:guardrails} as well as a series of optimizations and refinements that enhance its performance and accuracy~\sx{sec:optimizations}.
\end{itemize}

The paper then presents \sys's evaluation~\sx{sec:eval}, related work~\sx{sec:related-work}, and limitations~\sx{sec:limitations}.
An additional appendix contains the full set of prompts used (Appendix~\ref{appendix:prompts}).

% In the table, talk about whole domain like Program Synthesis, Example-Based Synthesis, etc.

\heading{Availability} 
\sys will be available as an MIT-licensed open source package for download at: 
\begin{center}
    \url{https://github.com/atlas-brown/lexo}
\end{center}

\section{The Event-Stream Attack}
\label{sec:example}

The \ttt{event-stream} library implements
stream manipulation functionality for \node~\cite{ev:eurosec:2022, eventStreamNPM}.
With over 3M weekly downloads,
\ttt{event-stream}
became the target of a supply-chain attack,
by introducing 
a Bitcoin-stealing malware
dependency called \ttt{flatmap-stream}~\cite{es1,es2,snyk-eventstream-2018,snyk-flatmapstream-2018}.

\heading{The malicious dependency} 
This malicious dependency extracts account details from Bitcoin wallets. 
The attack is stealthy: 
  it activates only when the malicious code executes as part of the Copay Bitcoin wallet~\cite{copay},
  on specific platforms, and 
  only when the wallet is connected to the live Bitcoin network~(lns. 3-4).
To perform the attack, \ttt{flatmap-stream}
(1) loads the user's wallet credentials,
(2) stores the credentials on the side, and 
(3) sends the credentials to a remote server~(code elided for simplicity),
before constructing and returning the \emph{correct} value to the caller method~(lns. 10-11).

\begin{minted}[xleftmargin=\parindent, linenos, numbersep=5pt, fontsize=\small]{js}
const Stream = require('stream').Stream;
module.exports = function (mapper) {
 if (env.Copay && env.livenet
  && env.platform === target) {
  // Code to (1) load wallet credentials
  // (2) store credentials on side, and
  // (3) send credentials to a remote server
}
const stream = new Stream();
// Construct flatmap stream object...
return stream; }
\end{minted}

\noindent
This is a SSCA,
as it does not
interfere with the component's client-visible behavior, even when activated.

\heading{Applying \sys}
% The way this attack was implemented makes it difficult to detect, 
% as it involved very specific activation conditions, sophisticated obfuscation techniques, and was embedded in a trusted package.
% Using \sys, a developer could generate a secure version of \ttt{flatmap-stream}, eliminating the dependency.
\sys regenerates a replacement \ttt{flatmap-stream} component
that provides identical client-visible behavior while
including no malicious code---and works across domains in a language-agnostic manner.
The following line invokes \sys on \ttt{flatmap-stream} with JavaScript as the output language.

\begin{verbatim}
$ lexo --out js flatmap-stream
\end{verbatim}

\noindent
\sys starts by 
extracting the interface and source code of the \ttt{flatmap-stream} component (step (1) in \cref{fig:system-overview}).
The system then continues with the \emph{candidate input generation} phase
(step (2) in \cref{fig:system-overview}), where it presents an LLM instance with
the \ttt{flatmap-stream} code and prompts the LLM to generate a set of inputs that
exercise the component's functionality. An example of a generated input is the JavaScript value \ttt{x => x + 1}. \sys then runs the \ttt{flatmap-stream} component on these 
generated inputs and collects the corresponding outputs to obtain a set of input-output examples.
To obtain a comprehensive set of input-output examples, \sys records the code coverage
that these input-output examples obtain, then iteratively prompts the LLM to generate additional
inputs to increase coverage. \sys currently performs three input generation iterations, stopping early
if the generated inputs achieve 100\% coverage. Note that if the component contains stealthy malicious code, the LLM typically does not obtain 100\% coverage because it does not construct the combination of inputs and execution environment required to activate the malicious code. 

In our example this process produces a total of 13 input-output pairs after three revisions, obtaining
% 73.91\% 
74\% code coverage (the input examples do not activate the malicious code). 
An example input-output pair is:
\begin{verbatim}
{ input: x => x + 1, 
  output: { new __Stream({ ... }) } }
\end{verbatim}
\noindent
\sys then moves on to the \emph{program regeneration} phase (step (3) in \cref{fig:system-overview}). 
Here, \sys presents the input-output pairs to the LLM, first prompting it to regenerate an algorithm sketch in natural language that implements the behavior specified by the input-output pairs, then next prompting it to refine the natural language sketch into a JavaScript component that implements the behavior (see \cref{sec:ablation} for results from experiments that go directly from the input-output pairs to source code without the intermediate algorithm sketch). 

%In practice this input-output-pair based regeneration approach produces regenerated programs that implement the behavior specified by the input-output pairs without side effects (such as file system or network access) orother extraneous code (input-output pairs cannot express these kinds of side effectsand LLMs typically do not insert this kind of code into theregenerated component unless prompted to do so). 

% XXX - can we also check for 100\% code coverage as a further way to 
% ensure no malicious code regeneration - 
% Vagos: Yes, I added this. I quickly tested it on 5 libraries and we get 100\% coverage on the regenerated code XXX

In the example, after three revisions this input-output-pair based 
regeneration approach produces the following code (shortened for brevity). This code implements the behavior specified by the 
input-output pairs without side effects (such as file system or network access) or
other extraneous code (input-output pairs cannot express these kinds of side effects
and LLMs typically do not insert this kind of code into the
regenerated component unless prompted to do so). 

\begin{minted}[xleftmargin=\parindent, linenos, numbersep=5pt, fontsize=\small]{js}
const Stream = require('stream').Stream;
const stream = new Stream();
stream.on('data', (x, f = v => v) =>
  (Array.isArray(x) ? x : [x])
    .forEach(y => out.emit('data', f(y))));
return stream;
\end{minted}

% \sys performs revisions after the candidate input generation and program regeneration phases. These are driven by a code coverage and correctness metric, respectively. Had code coverage not reached 100\%, \sys would have issued a warning and moved on to the program regeneration phase. If the program regeneration phase were to reach a maximum revision count while still failing on a configurable percentage of input-output examples (10\% by default), \sys would terminate the process with an error, stating that it is \emph{unable to regenerate the target library correctly}.

\heading{Key result}
The regenerated \ttt{flatmap-stream} component contains no malicious code, correctly implements all 13 input-output pairs used during regeneration, exhibits 100\% code coverage on this set of input-output pairs, and passes 5/5 (100\%) tests in the \ttt{event-stream} test suite~\sx{sec:eval-event-stream}.
Manual inspection confirms that the regenerated component 
correctly implements non-malicious \ttt{flatmap-stream} functionality.
The entire process takes a little over 10 minutes and costs \$0.10.
The initial input generation phase takes 30s
and incurs \$0.03 in LLM inference costs, 
followed by another 90s and \$0.07 for the program 
regeneration phase.

\begin{table}
  \centering
  \caption{
    \textbf{Comparison overview.}
    Approaches directly relevant to \sys's supply-chain security approach~\sx{sec:related-work}.
  }
  \label{tbl:system-comparison}
    \begin{adjustbox}{width=\columnwidth}
      \begin{tabular}{lcccc}
\toprule
Approach & \rotatebox{90}{\textbf{Generality}}
         & \rotatebox{90}{\textbf{Detection}}
         & \rotatebox{90}{\textbf{Elimination}}
         & \textbf{Examples} \\
\midrule
Static Analysis                    & $\squadfill$   & $\squadfill$  & $\squad$   & \cite{calzavara2015fine, maffeis2009language, xu2013jstill}  \\
Dynamic Analysis                   & $\squadfill$   & $\squadfill$  & $\squad$   & \cite{lya:fse:2021, dewald2010adsandbox, hu2018jsforce}  \\
Pure LLMs                          & $\squadfill$   & $\squadfill$    & $\squad$        & \cite{llmvulndetection2023, noever2023largelanguagemodelsfix, akuthota2023vulnerability}  \\
Programming-by-Example            & $\squad$        & $\squad$         & $\squadfillhr$ & \cite{jha2010oracle, feng2017component, Aspire_2015}  \\
Program Inference and Regeneration & $\squad$        & $\longrightarrow$ & $\squadfill$   & \cite{harp:ccs:2021, shen2019using, alr}  \\
\sys                               & $\squadfill$   & $\longrightarrow$ & $\squadfill$   & (this paper) \\
\bottomrule
\end{tabular}
\end{adjustbox}
\end{table}

\heading{Potential alternative approaches}
A security-conscious developer might try to proactively defend against this attack using
static or dynamic analysis,
% because key parts of the malicious payload are evaluated at run-time while being 
% encrypted at rest.
LLMs,
% as the malicious code activates when part of Copay wallet and only on production.
%inside a very specific environment%, far from development and testing. 
% When run in any other context, the compromised version of \ttt{flatmap-stream}
% exhibits identical behavior to the correct version.
or program synthesis. 
%Unfortunately, these techniques fail to successfully defend against the attack.

Static analysis~\cite{calzavara2015fine,Jamrozik2016,mirccs2021,Koved2002} is unable to detect the attack,
as the malicious payload is encrypted at rest and only decrypted and evaluated at run-time.
Dynamic analysis~\cite{lya:fse:2021, dewald2010adsandbox, hu2018jsforce} is also ineffective,
since the components's malicious side-effectful behavior activates only under strict conditions as described above.

LLMs have been used to directly detect and/or eliminate malware in component source code,
with the goal of producing a modified version of the component without malicious code
~\cite{alkaraki2024exploringllmsmalwaredetection, llmvulndetection2023, xu2024largelanguagemodelscyber}.
With a recent LLM (GPT-4o), this approach works for \ttt{flatmap-stream}, essentially by copying the legitimate code and discarding the malicious code. But given a modified version of 
\ttt{flatmap-stream} that includes additional safety checks and renames the payload to \ttt{util.js}, GPT-4o 
fails to recognize the malicious code and includes the 
malicious code in the modified version of the component. 
This example illustrates how this approach can fail, especially
when the attack is concealed through careful token selection or malicious code crafted (potentially automatically by an LLM, see \cref{sec:eval-naive}) to appear legitimate. 
\sys, in contrast, uses input-output pairs to sever any connection between the source code of the original 
and regenerated versions of the component, eliminating any need to detect and/or remove malicious code from existing code. 

Given a comprehensive enough set of input-output examples, traditional
program synthesis~\cite{harp:ccs:2021, feng2017component, Aspire_2015, chen2018synthesizing}
could, in principle, eliminate \ttt{flatmap-stream}'s malicious behavior by regenerating a behaviorally equivalent implementation.
However, this approach relies on domain expertise to develop a 
domain-specific language and corresponding specialized
synthesis engine for each target domain
(\eg string manipulation, array manipulation, \etc), which  requires programming language design and implementation expertise, can involve substantial engineering effort, and significantly
hinders scalability across domains.
It also requires the availability
of input-output examples that sufficiently characterize the desired component
behavior (standard component test suites, even when available, often do not contain enough examples). 

Tab.~\ref{tbl:system-comparison} summarizes the comparison, with additional points in the space of immediately relevant options~\cf{sec:related-work}. 

% (1): Provide to \sys the library -> we ask it to get some input (lets showcase some inputs)
% (2): We provide this inputs to potentially malicious library, (if something brakes we provide 
% it to the LLM again to generate better inputs for better code coverage) and we get the outputs 
% (lets showcase some of the outputs)
% (3): We give this inputs/outputs to the LLM to regenerate the library
% (loc)
% (4): And we get full passing suite on the event-stream (tests)

\section{Threat Model}
\vspace{-.1in}
\label{sec:threat-model}

% Full access to source
% Maintains benign behavior
% Adds malicious behavior
% But hides it, because otherwise it would be detected
% Activates only in target environments

\sys protects against attacks in which the attacker has full control of the target component and can modify its source code arbitrarily.
The modified target component exhibits correct client-facing behavior.
%(otherwise a malicious modification could be detected during testing, development, or production---\eg a crash can occur during the dependent application's execution).
The malicious behavior injected is side-effectful in that it produces effects 
beyond just producing component outputs, \eg filesystem access, environment
variable modification, network access, or arbitrary code execution.
% stealthy, in that it activates only in target environments or under precise conditions, and side-effectual, in that it produces effects beyond just producing the outputs of a function---
Since a malicious component can potentially attack the host system
during regeneration, \sys interacts with the component in an isolated
environment.

\sys also protects against software removal from remote repositories.
An adversary might unpublish a component
(as in the npm left-pad incident~\cite{leftpad}), replace it with a no-op, or replace it with an incorrect or 
incompatible version, thus breaking all dependent applications.
\sys can be deployed to proactively regenerate components
before removal, replacement, or revision, making alternative
versions available in the event of such removal, replacement, or revision.

% An assumption made is that a component's name is not purposely
% misleading. For example, a string padding library should not
% be named \ttt{math\_add} or \ttt{delete\_files}.

\sys assumes that the back-end LLM and the target language's runtime are all part of the trusted computing base.

\vspace{-.1in}
\section{System Overview}
\vspace{-.1in}
\label{sec:overview}

\begin{figure*}
    \centering
    \includegraphics[width=.8\textwidth]{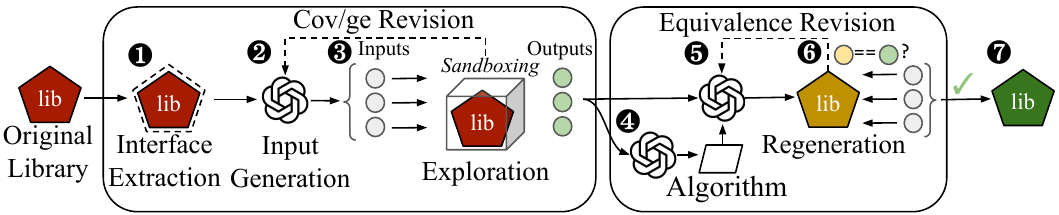}
    % https://docs.google.com/drawings/d/1kyC1aLbpqyvO-CP2BernFYhqczAwSMOOijBEso_6G88/edit?usp=sharing
    \caption{
      \textbf{\sys overview.}
% At a high level,
  \sys (1) analyzes a component's interface to extract its source, 
       (2) generates input-output pairs for the target component, (3) according to quality metrics,
       (4) extracts a new representation of the functionality informed by the input-output pairs,
       (5) regenerates the target component,
       (6) applies revisions until regenerated component conforms to the input-output pairs, and
       (7) replaces the original component with the regenerated one.
    }
    \label{fig:system-overview}
    \vspace{-.2in}
\end{figure*}

\sys takes as input a component's implementation
and produces a regenerated implementation that exhibits only the
component's client-facing functionality. 
This section explains the subsystems
used among different stages of the regeneration process
and how they combine together into the end-to-end pipeline.

\heading{Prompt anatomy}
\sys uses a set of prompts engineered to improve LLM 
generation performance~\cite{promptsurvey:arxiv:2024}. 
All prompts used in \sys adhere the following structure
(full prompts in \cref{appendix:prompts}):

\begin{enumerate}
    \item \textit{Problem statement}: An imperative explanation of the given task.
        For example:
        \begin{prompt-box-small}
        Given a component, generate a set of inputs that will thoroughly test the module's behavior. 
        \end{prompt-box-small}
    \item \textit{Problem break-down}: Elaborates steps to complete the task.
        For example:
        \begin{prompt-box-small}
1) Code Understanding: Explain the code's purpose and functionality.

\ldots

4) Explore: Include tests that are slight variations of the ones in the test suite.
        \end{prompt-box-small}
    \item \textit{Task clarifications}: In cases where it was observed that the LLM diverged from the desired output, the prompt includes some additional alignment instructions specifically mentioning the desired (or unwanted) patterns.
        For example:
        \begin{prompt-box-small}
        \ldots
        Make sure to test default arguments and optional parameters.
        \end{prompt-box-small}
    \item \textit{Output format specification}: 
        A short description of the desired output format.
        This enables parsing and utilization of the generated content
        by \sys.
        For example:
        \begin{prompt-box-small}
        Output one input per line, using types from the target language, in this format:
        [value1, value2]
        \ldots
        \end{prompt-box-small}
    \item \textit{Input data}: Data are provided to the LLM
        either as input to a transformation or additional context
        (\eg the component's source code, failed examples after a program synthesis attempt).
\end{enumerate}

% \sys's communication with an LLM is facilitated by the DSPy library~\cite{khattab2023dspy, khattab2022demonstrate},
% which is responsible for formatting prompts, 
% and interacting with model API endpoints.
% Because of the flimsy nature of the LLM's output,
% \sys also employs a set of additional guardrails to ensure
% that output follows an expected structure before it is processed by any of the
% subsequent modules.

\heading{Transformation modules}
The transformation modules  are responsible for translating a representation
of the target component from one form to another while preserving core
aspects of the original representation. 
In the current version of \sys, these are the component's input-output behavior and exported symbol name.
% XXX invariants from the original representation - how do invariants come in - XXX

% \begin{wrapfigure}{r}{0.5\columnwidth}
%     \hspace{-0.7cm}
%     \includegraphics[width=0.5\columnwidth]{figures/dspy_graph.pdf}
%     \caption{The DSPy~\cite{khattab2023dspy} graph \sys implements.}
%     \vspace{-0.3cm}
%     \label{fig:dspy-graph}
% \end{wrapfigure}

To initiate component learning, the \emph{inference} module, given the source component's implementation, generates test inputs for
the source component (step
(2) in \cref{fig:system-overview}). 
These inputs use only constructs and syntax from the source language. 
To obtain corresponding outputs, each generated input is executed on the
target component.
These input-output pairs capture only pure computations between the target component's
input and output.
% a wide range of behaviors commonly seen in third-party components, yet strict enough to omit any side-effectful or malicious behavior, striking
% a balance between generalizability and security.
Subsequent transformation modules leverage this property as a firewall,
eliminating attacks embedded in the original source component as
side-effectful computations cannot be represented.

To enhance interlocks between the generated input-output pairs and the target
component, the \emph{input-output pairs to algorithm} module utilizes the generated input-output
pairs to craft a high-level natural language algorithm.
This algorithm outlines
a general approach that, if implemented, would adhere to the given examples
(step (4) in \cref{fig:system-overview}). This stage of the pipeline prompts
the LLM to analyze the program synthesis problem, deriving
high-level insights from the examples before proceeding to
implementation. %---mimicking the way a human might approach this task.

\sys's final transformation is
the \emph{input-output pairs and algorithm to code} module.
The module takes as input 
(1) the source component input-output pair examples,
(2) the natural language algorithm generated by the input-output pairs to algorithm module,
and
(3) the source component's published name.
Using these inputs, \sys regenerates the component.
(step  (5) in \cref{fig:system-overview}).
% This transformation additionally utilizes the name of the original module,
% hinting to the LLM the target component's intended functionality.
% This can be especially potent in
% cases where the component's name alludes clearly to its functionality, as in
% libraries like \ttt{cartesian-product} or \ttt{camel-case}. 

\heading{Completeness module} \sys employs a set of metrics that quantify several
aspects of the input-output pairs and regenerated components
(steps 3 and 6 in \cref{fig:system-overview}). In the input-output pair generation
module, the module targets a
code coverage threshold (100\%) on the source component.
In the code regeneration module, the module verifies that the regenerated component complies with the specifications of the generated input-output pair set. 
Each produced artifact is considered correct if it passes all the correctness criteria.
If the input-output examples fail to achieve 100\% code coverage on the original component, \sys proceeds regardless. 
This accounts for the scenario of stealthy, malicious code that does not execute during testing.

\heading{Revision module}
In cases where accuracy or completeness thresholds are not met,
\sys's \emph{revision} module activates and attempts to improve the quality of
the generated content.
The revision module operates in a closed loop, with the LLM's output from one
iteration feeding back as part of the input for the next one.
\sys performs a tuneable number of revisions (three by default), after
which, if the correctness module has not marked the output as correct, the
system stops and discards the generated content, issuing an error message.
Otherwise, the output is saved on disk and \sys continues execution.

\vspace{-.1in}
\section{Input-Output Generation}
\vspace{-.1in}
\label{sec:io}

\heading{Input generation}
Given a component implementation $C$, \sys generates a set of inputs which is
used to construct a model of $C$'s client-observable 
behavior. 
The system loads the component and extracts from it each function $f_{n}$ and its corresponding source code $\text{src}(f_{n})$.
It then provides $\text{src}(f_{n})$ to
an LLM instance, instructing it to generate a set of inputs $I$ that exercise $f_{n}$.
\sys's regeneration happens on a per-function basis.
Each input
$i$ contains the values to be passed as inputs to $f_{n}$.
To maintain compatibility with later stages of the pipeline,
\sys uses a constrained input format that resembles
the target component's language syntax.
The format is designed to be easy to parse and convert into runtime values of the component's language.
Given a source language $\mathcal{L}$,
the input representation is an array in $\mathcal{L}$.
Each element of this array must be a valid expression in $\mathcal{L}$ and can be of any primitive or non-primitive type in $\mathcal{L}$, including function-like constructs.
Comments, as defined by the syntax of $\mathcal{L}$, may accompany the array, but are discarded, as they might leak a description of potentially malicious behavior.
The order of the elements within the array is significant and must be preserved as part of the input specification.
Multiple such arrays can be provided, delimited by a character or sequence of characters.

\sys provides the LLM with the input format's syntax and semantics for the source language as part of 
the input generation prompt using three examples~(\cref{appendix:prompts}).
% LLMs have shown high performance in generating text using specific formatting instructions~\cite{xia2024fofobenchmarkevaluatellms}.
Some examples of generated inputs are shown below:
\begin{prompt-box-small}
\begin{Verbatim}[fontsize=\small]
[{a:1,b:2}, "a"] // An input with two arguments
[1,3] // An input with two integer arguments
["hello"] // An input with a string argument
[(a,b)=>{a+b}] // An input with a function
\end{Verbatim}
\end{prompt-box-small}

While constrained decoding~\cite{park2024grammaraligneddecoding,
willard2023efficientguidedgenerationlarge} can be used to guarantee inputs adhere to a specific grammar,
\sys does not leverage this feature,
because of (1) possible regeneration performance degradation~\cite{tam2024letspeakfreelystudy},
(2) \sys's use of prompting techniques like chain-of-thought~\cite{wei2023chainofthought} that require 
the LLM to freely produce text in a scratchpad-like manner before giving a final answer,
and (3) compatibility with most LLM API services, as not all support constrained decoding.
% Empirically, we report that input generation never failed because of inability from the LLM to comply
% with the input value format.

% \sys uses a format that resembles JSON due to its simplicity
% and the fact that
% many commercial large language models (LLMs) have been trained on
% extensive datasets that include JSON,
% leading to better generation performance in this format~\cite{xia2024fofobenchmarkevaluatellms}.

It is important that the input-output pairs generated capture enough 
of the original component's legitimate behavior to enable correct regeneration.
Towards this goal, 
\sys instructs the LLM to generate both nominal inputs
that exercise the target component's typical behavior
and edge-case inputs that test the component's behavior in unusual or boundary scenarios.
\sys can deduce input argument types
even in the absence of developer type annotations,
inferred by the back-end LLM based on the operations
performed on them.
Even in cases where only the component's interface (and not source code) 
is available (as is the case
for native libraries), \sys can generate quality inputs informed by context
such as the function and argument names.
Combining the above, \sys can often reach all of the function's logic with only
a small set of inputs~(\cref{tbl:eval-alllibs}).

\heading{Input filtering}
After \sys instruments the LLM instance to generate a set of inputs $I$ for $f_{n}$,
it parses and splits the generated set into individual inputs $i$.
The system then inspects each generated input for syntactic correctness
using the source language's parser. \sys rejects any invalid inputs
and keeps a record of all valid inputs between revisions~(\cref{sec:guardrails}), adding new ones to the set
at each iteration.

% \begin{figure}[h]
% \centering
% \[
%   \begin{array}{lrcl}
%   \text{Test Case} & \mathit{T} &::=& \{ \mathit{A} \} [ // \mathit{D} ] \\
%   \text{Arguments} & \mathit{A} &::=& \mathit{P} ( , \mathit{P} )^* \\
%   \text{Argument Pair} & \mathit{P} &::=& \mathit{N}: \mathit{V} \\
%   \text{Argument Name} & \mathit{N} &::=& [a-zA-Z\_][a-zA-Z0-9\_]* \\
%   \text{Value} & \mathit{V} &::=& \mathit{X} \vv \mathit{L} \vv \mathit{O} \\
%   \text{Primitive} & \mathit{X} &::=& \mathit{S} \vv \mathit{N} \vv \mathit{B} \vv \mathbf{null} \\
%   \text{Array} & \mathit{L} &::=& [ \mathit{V} ( , \mathit{V} )^* ] \\
%   \text{Object} & \mathit{O} &::=& \{ \mathit{P} ( , \mathit{P} )^* \} \\
%   \text{Description} & \mathit{D} &::=& [a-zA-Z0-9\_ ]+ \\
%   \text{String} & \mathit{S} &::=& " [^"]* " \\
%   \text{Number} & \mathit{N} &::=& [0-9]+ ( "." [0-9]+)? \\
%   \text{Boolean} & \mathit{B} &::=& \mathbf{true} \vv \mathbf{false} \\
%   \end{array}
% \]
% \vspace{-12pt}
% \caption{
% \textbf{The input format \sys uses to generate input-output pairs.}
% }
% \label{fig:input_format}
% \end{figure}

\heading{Output capture}
After input generation, each input is serialized
back into the source language's syntax.
Then, \sys creates a program written in the target language that includes
a routine which executes each one of the generated inputs $i$ against the original component.

\heading{Input/Output filtering}
At this point, \sys
takes each input $i$ and 
provides it to the original function $f_{n}$,
resulting in a corresponding output $o$.
\sys dynamically adds serialization logic and records the component's output from standard out.
Interactions with the component happen in an isolated environment to prevent any side effects from affecting the host system.
But beyond side-effects, the component's output $o$ may include values that may pollute later stages of the pipeline.
For instance, a component that implements padding functionality may be invoked with a very large padding length, for example \ttt{pad("s", 999)}, polluting the context window with a long string of spaces.
% This process may sometimes produces very long outputs, which may pollute or overflow the LLM's context window during later stages of the pipeline.
To protect against
such context window exhaustion or overflow, \sys removes input-output pairs that have combined length over a tuneable threshold (100 tokens by default).
After this, the system inspects the output for correctness in the context of the source language
and, again, serializes it.
\sys records the component's behavior both in cases of valid output returned as well as any exceptions thrown,
in which case it records the exception's type and message.
\sys then adds each input-output pair tuple $\langle i, o \rangle$
into its set of example input-output pairs $E$.

\vspace{-.1in}
\section{Regeneration}
\vspace{-.1in}
\label{sec:synthesis}

After generating a set of input-output pairs that adequately capture 
the $f_{n}$ client-observable behavior,
\sys moves on to regenerate a replacement function $f_{n}'$.

\heading{Natural language algorithm}
\sys starts by inspecting the generated input-output pair set $I$
and instructs the LLM  to produce an algorithm in natural language 
that describes at a high level the implementation of a function
that would satisfy the input-output pairs (full prompt in \cref{appendix:prompts}).
% $f_{n}(i) = f_{n}'(i)$ for all $i \in I$.
\begin{prompt-box-small}
Given this test suite, design an algorithm that describes the function.

1) Understand the tests

2) Analyze the problem

3) Design the algorithm

4) Handle edge cases
\end{prompt-box-small}

\noindent
A snippet of an algorithm generated for the \ttt{flatmap-stream} library is:

\begin{prompt-box-small}
1. Check if the input is `undefined`. If so, return `[]`.
2. If the input is not an array, iterate over each element and apply the mapping function.
\ldots
\end{prompt-box-small}

\sys then provides this natural language description to the LLM along with the set
$E$ of input-output pairs.
Before providing input-output pairs to the LLM, \sys first converts the intermediate input-output pair format 
into a form that more closely resembles the syntax of the target language.
This helps instruct the LLM to synthesize the target component using input-output pairs, without 
having to re-state the semantics of the chosen input-output pair format.
For example, the input-output pair \ttt{\{input: [3, 4], output: 7\}} for a component \ttt{f} that implements addition
would be
included in the prompt as \ttt{f(3, 4) = 7}.
\sys includes input-output pairs that exercise incorrect input types or
values that result in exceptions, 
as input sanitization and error conditions
from the original library should be retained after regeneration.

% The program synthesis prompt includes all generated input-output pairs.

\heading{Equality checks}
After each synthesis attempt, \sys verifies that \( f_{n}'(i) = o \) holds for every input-output pair \(\langle i, o \rangle \in E\).
\sys uses the language's \emph{deep-equality} operation, which recursively compares the two values.
The operation traverses language values down to their primitive types, where it then checks for literal equality.
% In the case of function-like values, \sys performs pointer equality.
% Maybe talk about pointer equality

% For example, consider the identity function $f_{I}$.
% During \sys's input generation phase, 
% one of the generated inputs might be a function like \ttt{l = (a, b) => { a + b }}.
% Even if \sys had synthesized a correct implementation $f_{I}'$, 
% $f_{I}'(l) \neq f_{I}(l)$, since most programming languages
% To circumvent cases \sys serializes both 
% values and compares their string representation, 
% a more lenient form of equality.

\vspace{-.2in}
\section{Guardrails}
\vspace{-.1in}
\label{sec:guardrails}

% \sys applies guardrails to reject both malformed input candidates and incorrect synthesized programs.

During candidate input generation, \sys checks if the final input set
results in 100\% code coverage of the original component.
If not, it enters a revision loop, in which it generates additional inputs
using code coverage as a metric.
In this loop, 
the LLM is instructed to focus on edge cases present in the given source code.
If the code coverage condition threshold fails,
the input-output generation part of the implemented pipeline is re-run, now modified,
with the failed 
generation being appended to the instruction prompt, together with a message
instructing the LLM to create a broader range of inputs to increase coverage.
After three revisions, \sys moves on to the program regeneration part of the implemented pipeline,
using the input-output pairs it has generated so far.
Reaching 100\% code coverage might not be possible, as parts of the code could be intentionally unreachable or activated only in specific environments as part of an attack.

During program regeneration \sys checks if the current regenerated version correctly implements the generated input-output pairs. If not, 
\sys reruns the program synthesis step, modifying the process
by appending the current version to the LLM instruction prompt. It also
includes a message prompting the LLM to revise the code so that it correctly implements all input-output pairs, along with the specific examples that caused the
failure.
After three revisions, if the correctness module has not marked the output as correct,
\sys stops and discards the synthesized code, issuing a warning message to the user.

After regeneration completes, \sys executes any tests created by the
original developers (if available) and issues an error 
if the regenerated component does not achieve 100\% correctness.
% To verify that no side-effectful behavior was
% introduced during the regeneration process,
% \sys runs the regenerated component against the previously generated inputs in a sandboxed environment, recording any file-system acceses using the 
% \ttt{try}~\footnote{\url{https://github.com/binpash/try}}
% utility.
% If any file-system access is observed, \sys exits and produces an error,
% as it wasn't able to eliminate the component's side-effectful (and possibly malicious) behavior.

% Make a wrapfigure
% \begin{wrapfigure}{r}{.5\columnwidth}
%    \hspace{-.7cm}
%    % https://docs.google.com/drawings/%d/1RMn_Hbgjxs6wzU1GMqvsevgu1Yc991MQxbzm30-wmos/edit?usp=sharing
%    \includegraphics[width=.5\columnwidth]{figs/tainted.pdf}
%    \caption{Before the source component is converted into input-output %pairs, \sys treats it as tainted data.}
%    \label{fig:tainted-input}
%    \vspace{-.3cm}
%\end{wrapfigure}

To avoid malicious code regeneration,
the original  implementation is treated as ``tainted''.
It is then turned safe by projecting it to a set of input-output pairs that capture its behavior.
When \sys prompts the LLM to regenerate a new version, the
LLM is not given the original potentially malicious code as input. 

%to analyze the synthesis problem~(\cref{sec:prompt-techniques}), it does not use the original implementation directly as input, as the LLM might consider the malicious snippet as part of the original component, include it in its analysis, and regenerate it. Thus, subsequent LLM instances \sys instruments are provided only with the input-output pair representation of the component.

% % This might be moved in the introduction
% A driving factor to the development of our \textit{prompt transcryption} technique were interlocks.
% These are cases where when malicious code is considered by the LLM as being part of the original component.
% If that is the case, then the naive approach of directly asking the LLM whether a piece of code contains malicious snippets fails.

% While \sys does not provide absolute guarantees, it not only decouples the
% regenerated code from the original source code but also introduces necessary
% safeguards. This approach leverages the strengths of general-purpose Large
% Language Models (LLMs) in code generation,
% while absolving the model form the need to understand that a given code snippet is malicious.

\vspace{-.1in}
\section{Optimizations and Refinements}
\vspace{-.1in}
\label{sec:optimizations}
%Heavily referenced the Medprompt paper for this section: https://arxiv.org/pdf/2311.16452.pdf

%This section presents several optimizations and refinements. 
% that \sys employs and refinements that have been made to the system.

% \subsection{Prompting Techniques}
\label{sec:prompt-techniques}
% \sys uses a number of prompting techniques to improve generation quality.

\heading{Chain-of-thought prompting technique}
The chain-of-thought (CoT)~\cite{wei2023chainofthought} prompting technique
explicitly encourages a language model to generate a rationale before answering.
This approach has been shown to improve the ability of foundation models
to perform complex tasks~\cite{feng2024towards, li2023structured, saparov2022language, wu2023chain}.
\sys uses a CoT stage for each LLM-assisted task.

\heading{Deconstruction}
During the code regeneration phase,
\sys deconstructs the problem into smaller sub-tasks, a technique that
has been shown to improve the ability of foundation models to
perform complex reasoning \cite{khot2023decomposed}.
Before regeneration, \sys uses the generated set of input-output pairs to instruct the LLM
to create a natural language description of the final
program's functionality.
This allows the LLM to sketch an initial plan \cite{nye2021work}, where it pieces together, based on the given examples, what the final module's functionality should be.
Then, \sys gives this high-level description as part of the regeneration
prompt alongside the generated input-output examples, when \sys prompts the LLM to
regenerate the component.
Intuitively, this helps the LLM 
draw from this high-level description of the program's functionality alongside concrete inputs and outputs.

% % Maybe add an example here? I feel like this is important for the reader to understand. Consider function-compose

% \subsection{Parallelization}
\heading{Parallelization}
\sys performs regenerations in an isolated fashion 
and is able to regenerate multiple components in parallel.
The parallelization boundary is a single component.
When run with the \ttt{-{}-parallel} flag, \sys begins regenerating the specified set of software components in parallel using half of the system's available cores. 
\sys schedules regeneration using lines-of-code as a priority metric,
starting with smaller libraries first.

\heading{Language Agnosticism}
\sys achieves language agnosticism both in the fact that it leverages LLMs to infer a component's functionality and regenerate it,
but also in the way it interacts with the target component.
Specifically, \sys interacts with the target component in a language-agnostic manner during input-output generation,
as it only needs to provide inputs and observe corresponding outputs.
To support a new language, \sys requires a minimal language-aware component, which is responsible for loading the target component, extracting its source, and executing it with generated inputs while capturing outputs and line coverage.

% \subsection{Early Exits}
\heading{Early exits}
When invoked with the \ttt{-e} flag, \sys will inspect the target component to identify 
if it falls under its model of regeneratable components.
During input-output generation, when \sys exercises the component's behavior on the set of generated inputs, 
if \sys observes that one of the component's return values are outside of \sys's regeneration capabilities,
\sys will abort regeneration and issue an error.
Specifically,
\sys aborts when the return value of a component is of type \ttt{Function}, as
during verification it can not attest equivalence between two function-like values.
The system will also abort in case one of the LLM-generated inputs or component-produced 
outputs are not serializable.
\sys also discards components that, 
access the system's time, access the filesystem or make network requests by running them through its generated input-output pairs while monitoring them using \ttt{strace}~\cite{strace}. %, and leveraging the tool's classification of system calls as ones that interact with the file-system (\ttt{\%file} set, or network.
To avoid false negatives coming from the language runtime's own system calls,
\sys first performs a pure run with a dummy component inside the given runtime and filters out any system calls that appear during this run from the final monitoring step. 

\section{Evaluation}
\label{sec:eval}

% \begin{figure*}[t]
%     \centering
%     \includegraphics[width=\textwidth]{figures/correctness.pdf}
%     \caption{\textbf{Results of component regeneration correctness.} 
%     $x$-axis represents libraries from npm.
%     $y$-axis measures the percentage of developer-provided tests successfully passed by the regenerated libraries.
%     Shading represents the percentage of input-output pairs that the regenerated component passes.
%     Note that \sys automatically reports inability to regenerate when correctness is below 100\%.
%     }
% \end{figure*}

\begin{figure*}[t]
    \centering
    % Subfigure 1: GPT-4.0
    \begin{subfigure}[t]{\textwidth}
        \centering
        \includegraphics[width=\textwidth]{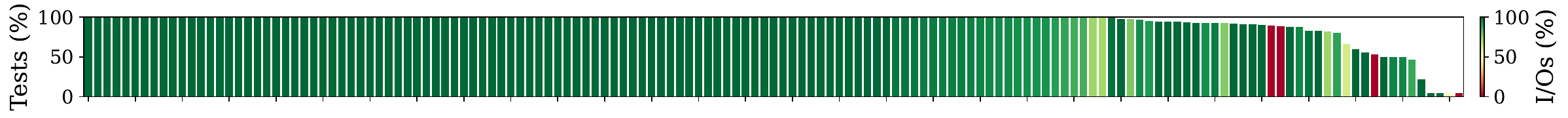}
        \caption{\textbf{Correctness results for \sys using GPT-5 mini~\cite{gpt5systemcard}.}}
    \end{subfigure}
    \begin{subfigure}[t]{\textwidth}
        \centering
        \includegraphics[width=\textwidth]{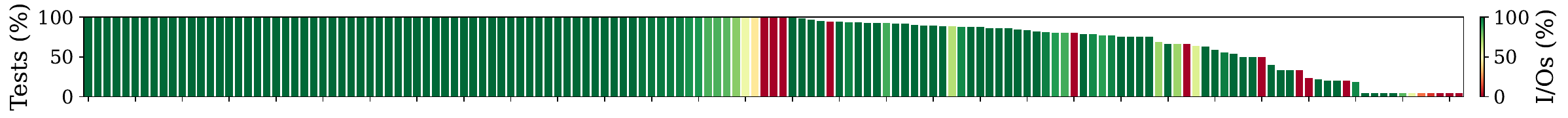}
        \caption{\textbf{Correctness results for \sys using GPT-4o~\cite{openai2024gpt4ocard}.}}
    \end{subfigure}
    % Subfigure 2: GPT-3.5
    \begin{subfigure}[t]{\textwidth}
        \centering
        \includegraphics[width=\textwidth]{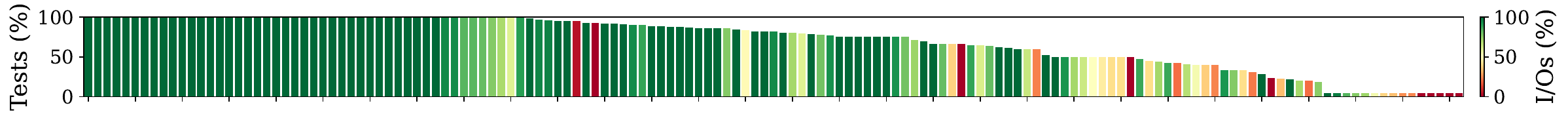}
        \caption{\textbf{Correctness results for \sys using GPT-3.5~\cite{ye2023comprehensivecapabilityanalysisgpt3}.}}
    \end{subfigure}
    % Subfigure 3: Mistral
    \begin{subfigure}[t]{\textwidth}
        \centering
        \includegraphics[width=\textwidth]{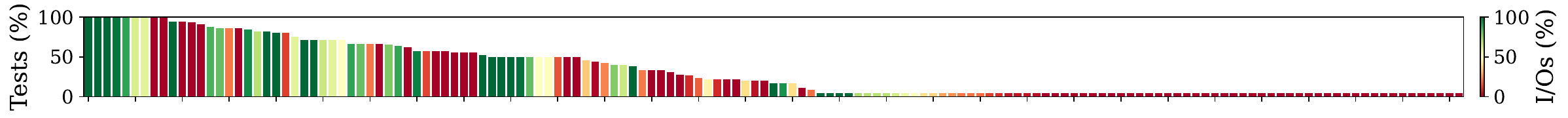}
        \caption{\textbf{Correctness results for \sys using Mistral~\cite{jiang2023mistral7b}.}}
    \end{subfigure}

    \caption{\textbf{Regeneration correctness results across models.} 
    Displaying the results of applying \sys on {\totalLibs} libraries from open source component ecosystems.
    Each bar represents a different component regenerated by \sys.
    The $y$-axis represents the percentage of developer-provided tests successfully passed by the regenerated libraries.
    Shading represents the percentage of input-output pairs that the regenerated component passes.
    \sys automatically reports inability to regenerate when correctness is below 100\%.
    The same set of libraries is present in all three plots in descending order of correctness.}
    \vspace{-.1in}
\end{figure*}

This section presents results from applying \sys to real-world components to answer the following questions:

\begin{itemize}[leftmargin=*, nosep]
  
  \item \textbf{Q1:} How effective is \sys at eliminating vulnerabilities in real-world malicious components?
  \sys successfully eliminates vulnerabilities that enable 
  ten large-scale software supply-chain attacks.
  \sx{sec:eval-security}

  \item \textbf{Q2:} To what extent can \sys correctly learn and regenerate
      client-facing component behavior? 
    \sys fully regenerates \sysCorrectLibs/\totalLibs libraries (100\% of developer tests passing, followed by manual inspection; all components that correctly implement the generated input-output pairs and pass the developer tests also pass manual inspection).
  \sx{sec:correctness}

  \item \textbf{Q3:} How efficient and scalable is \sys's regeneration when applied to
      real-world components? 
  Regenerated components have an average performance overhead of less than 0.1\% over the original implementations.
  \sx{sec:performance} 

  \item \textbf{Q4:} How does \sys compare to using an LLM to recognize and eliminate malicious code?
  \sys outperforms this approach 
  in eliminating malicious code from all components implemented as native
  NodeJS addons, and the synthetic \ttt{leetlog} and \ttt{pad}
  supply-chain attacks.
  \sx{sec:eval-naive} 

  \item \textbf{Q5:} How does \sys compare to state-of-the-art domain-specific program inference and regeneration techniques within the target domain?
      Applied to the 17 components from the Harp benchmark set~\cite{harp:ccs:2021}, 
  Harp's domain-specific synthesis engine and accompanying DSL  correctly regenerates 14/17 components while \sys correctly regenerates 
  5/17 components.
  \sx{sec:program-synthesis}
\end{itemize}

\heading{Software and hardware setup}
All experiments were conducted on a server with 
62GB of memory and 
12 physical cores with an
2.20 GHz Intel Core(TM) i7-10710U CPU @ 1.10GHz CPU,
running Ubuntu 24.04.1 LTS 6.8.0-41-generic. 
The setup uses Node.js v18.19.1 and CPython 3.12.3.

\heading{Implementation details}
The \sys prototype core is
developed in 1283 lines of Python code, and is complemented by three language-specific libraries for interfacing with Python, JavaScript, and Ruby components~(approximately 100 lines each).
\sys uses the \ttt{dspy} library~\cite{khattab2023dspy}, a library that provides a high-level interface for interacting with LLMs using a constrained programming model.
\sys's two stages are implemented as two \ttt{dspy} modules: one for generating inputs and obtaining input-output pairs, and another for synthesizing the component from the input-output pairs.

\begin{table*}[t]
  \centering
  \caption{\textbf{Summary of correctness and performance results from applying \sys on open source libraries.}
  Each row presents the library's name;
  target domain ($\mathcal{D}$)---(A)rrays, (S)trings, (F)unctions, (O)bjects, (C)ollections, (H)as, (I)s, and (M)ath);
  source language ($\mathcal{L}$);
  number of weekly downloads and number of dependent packages;
  performance of the input-output pair generation phase across runtime ($t_{gen}$), number of revisions ($R_{i}$), number of generated input-output pairs (P), and code coverage (Covg);
  and performance of the regeneration phase across runtime ($t_{synth}$), number of revisions ($R_{s}$), number of input-output pairs correctly implemented (Correct$_i$), number of developer tests passed (Correct$_d$), and performance overhead (Perf/nce) compared to the original implementation.
  For these results, \sys uses \gptmodel as the back-end LLM.
  }
   \label{tbl:eval-alllibs}
  \footnotesize
% \multicolumn{3}{c}{Library}   & \multicolumn{2}{c}{Popularity} & \multicolumn{4}{c}{I/O Pair Generation} & \multicolumn{4}{c}{Program Synthesis} & \\
% Name & $\mathcal{D}$ & $\mathcal{L}$ & Weekly & Deps & $t_{gen}$ (s) & $R_{i}$ & P & Covg & $t_{synth}$ (s) & $R_{s}$ & Correct$_i$ & Correct$_d$ & Perf/nce \\
  \input{table.tex}
\end{table*}

\vspace{-.1in}
\subsection{Security}
\vspace{-.05in}
\label{sec:eval-security}

To answer \textbf{Q1}, we collected ten libraries
from publicized supply-chain security incidents~\cite{ev:eurosec:2022, leftpad},
the Backstabber's Knife Collection (BKC)~\cite{ohm2020backstabber},
and hand-crafted examples designed to by bypass 
an LLM's ability to detect malicious code.
To confirm that \sys successfully eliminates each attack
we manually inspected the regenerated libraries
to ensure that \sys produces a secure and vulnerability-free implementation,
without introducing new vulnerabilities.
% and (2) applied the \ttt{mir-sa} tool~\cite{vasilakis2021mirautomatedquantifiableprivilege} to the regenerated libraries
% and confirmed that no calls to sensitive APIs (\ie \ttt{eval}, \ttt{require}, \ttt{child\_process.exec}, \etc) were present.
The following paragraphs present the results of applying \sys to eliminate malicious or vulnerable code.

\heading{Event-Stream}
\label{sec:eval-event-stream}
The \ttt{event-stream} incident~\cite{ev:eurosec:2022} 
is a supply-chain attack that introduced a malicious dependency into the \ttt{event-stream} component.
The setup for the \ttt{event-stream} component includes the original component~\cite{tarr2018eventstream}
and a re-implementation of the malicious \ttt{flatmap-stream} dependency~\cite{hugeglass2024flatmapstream}.
The incident is discussed extensively in \cref{sec:example}.
\sys successfully regenerates \ttt{flatmap-stream},
removing the activation conditions checks %(code verifying the component is running inside the Copay wallet, is on the livenet, \etc) 
and the payloads themselves.

%\heading{Firefox}
%\label{sec:eval-firefox}
%The \ttt{Skia} graphics component used by Mozilla Firefox had a %critical 
%buffer overflow vulnerability~\cite{skiaVulnerabilityGithub}. 
%%This vulnerability occurred because
%The \ttt{Skia} component performed improper 
%buffer offset calculations using 32-bit arithmetic, 
%which led to potential overflow conditions during hardware-accelerated 2D canvas actions.
%This vulnerability allowed for a potentially exploitable crash that could be 
%leveraged for remote code execution or denial of service attacks.
%We inject the same vulnerability to a JavaScript character-counting component, implemented 
%as a native Node.js addon, and use \sys to regenerate it.
%\sys successfully regenerates the component, removing the buffer overflow 
%vulnerability, producing memory-safe JavaScript code that does not use memory-unsafe language features.

\heading{CTX}
The \ttt{ctx} PyPi component implements a Python dictionary wrapper that allows items to be read using attribute access notation.
The component's \ttt{v0.2.6} release extended the exported \ttt{class} with a method called \ttt{sendRequest}, that sends all environment variables to a remote server~\cite{Sharma2022ctx}.
\sys successfully regenerates each method in the component, and turns the \ttt{sendRequest} method into a no-op, removing the exfiltration behavior.

\heading{Strong-password}
The \ttt{strong-password} component is a Ruby library that performs password strength analysis.
On the component's \ttt{v0.0.7} release, a malicious snippet was introduced that defines a self-executing method that executes a script fetched from a remote server~\cite{cve:strong-password}.
The attack only activates when it detects that it is running in a production server using the Ruby on Rails framework.
Applying \sys on the malicious function discards the malicious behavior, regenerating it as a function returning \ttt{nil}.

\heading{Passgen}
Similar to \ttt{strong-password}, the \ttt{passgen} library is a Ruby component that generates random passwords based on a set of rules.
On the component's \ttt{v1.0.3} release, an almost identical malicious snippet to the one in \ttt{strong-password} was introduced~\cite{ohm2020backstabber}.
Applying \sys on the exported functions inside the \ttt{probabilities.rb} file results in an implementation that correctly exports the probability matrix used to generate passwords and turns the malicious method into a method that returns \ttt{nil}.

\heading{Left-Pad}
\label{sec:eval-left-pad}
The \ttt{left-pad} incident~\cite{Quartz2016} occurred when a popular JavaScript 
library,
used by multiple projects such as 
React and Babel, was unpublished,
affecting production 
environments such as LinkedIn and Facebook.
\sys is applied to the component built by npm in response to the incident. 
\sys successfully regenerates the component,
which when used as a drop-in replacement passes all developer-provided test cases,
with a performance overhead of 1.1\%.

\heading{String-Compare}
\label{sec:eval-string-compare}
%https://naildrivin5.com/blog/2019/07/10/the-frightening-state-security-around-npm-package-management.html from harp paper
The \texttt{string-compare} attack involves the coexistence of
two versions of a single component within the same codebase~\cite{Foy_2019}.
One version is harmless and is used in a sorting function.
The other version, used as a dependency of an authentication module,
exhibits malicious behavior when it receives the specific
authentication string \texttt{gbabWhaRQ},
at which point it gains access to the file system of the server
on which the program is operating.
\sys successfully regenerates both components,
removing the malicious code from the second version.

\heading{Leetlog}
\label{sec:eval-leetlog}
The \ttt{leetlog} component provides functionality to log strings to the console
in a variety of different colors.
The component contains an actively malicious snippet 
that appends a predefined SSH key to the authorized keys to a target user of the system~\cite{github_advisory_GHSA-gfm8-g3vm-53jh}.
\sys successfully regenerates the component, eliminating the malicious code.

\heading{Synthetic pad}
\label{sec:eval-pad}
The \texttt{pad} component is a
string padding function that contains a Remote Code Execution (RCE) vulnerability.
The component, apart from its benign functionality, contains
a section that fetches a code snippet from a remote server and executes
it by converting the fetched code into a JavaScript function object and calling it.
\sys successfully regenerates the component and removes the vulnerability,
synthesizing only the padding functionality,
as the fetching of the URL constitutes side-effectful behavior.

\heading{Synthetic left-pad}
\label{sec:eval-synthetic-left-pad}
The \texttt{synthetic-left-pad} component is a modified version of the \texttt{left-pad} component~\cite{leftpad}.
The component performs its intended functionality as expected,
except for an attack that activates if the \texttt{PRODUCTION} environment variable is set to \texttt{true}.
In that case, the component accesses the file system, reading a file specified by the input
and logging its contents to the console.
The component includes checks for known JavaScript vulnerabilities, such as path traversal,
and ostensibly follows best practices by using try-catch blocks and throwing errors when it deems the input to be unsafe.
\sys successfully regenerates component's padding functionality without the malicious filesystem access.

\heading{Benign components}
\label{sec:eval-beign}
\sys is also applied to \totalLibs benign components
that feature no known vulnerabilities or malicious behavior.
After manually inspecting each regenerated component
and applying the \ttt{mir-sa} static analysis tool to extract 
sensitive API calls,
we find that \sys does not introduce new vulnerabilities or malicious behavior,
even in cases where it does not reach 100\% correctness.

\subsection{Correctness}
\label{sec:correctness}

For \textbf{Q2}, we apply \sys on a collection of 
\totalLibs benign components spanning 7 different domains~(\ttt{array},
\ttt{collection},
\ttt{is},
\ttt{math},
\ttt{object},
\ttt{string}, 
\ttt{function})
from the npm, PyPi, and RubyGems ecosystems.
We also include native modules written in C and C++, loaded into Node.js via \ttt{require}.
% Components used are 4--230 lines of code, averaging 32 lines of code.
To confirm correctness of the regenerated components,
we (1) use developer-provided test-cases 
and (2) manually inspect the regenerated component's source code.

We compared \sys's performance across four models of
varying sizes and capabilities over the \totalLibs components:
Mistral 7B (7B parameters)~\cite{jiang2023mistral7b},
GPT-3.5 (175B parameters)~\cite{ye2023comprehensivecapabilityanalysisgpt3},
GPT-4o (1T+ parameters)~\cite{openai2024gpt4ocard},
and GPT-5 mini (1T+ parameters)~\cite{gpt5systemcard}.

Results show that with Mistral, \sys exhibits the weakest performance: 3 of the \totalLibs regenerated components passed both input-output pairs and developer tests,
15 passed input-output pairs but not developer tests, and the rest did not pass input-output pairs.
With GPT-3.5, 
35 passed both input-output pairs and developer tests,
14 passed input-output pairs but not developer tests, and the rest did not pass input-output pairs.
With GPT-4o,
59 passed both input-output pairs and developer tests,
43 passed input-output pairs but not developer tests, and the rest did not pass input-output pairs.
With GPT-5 mini,
\sysCorrectLibs passed both input-output pairs and developer tests,
18 passed input-output pairs but not developer tests, and the rest did not pass input-output pairs.
Manual inspection indicates that, for
all LLMs, all versions that passed both input-output pairs and developer tests correctly implement the intended component functionality.
None of the generated versions, whether correct or not, contained malicious code.  

An examination of the synthesized programs reveals 
additional information. 
\sys-generated code often lacks the optimizations commonly
implemented by developers, such as early exits or caching.
For instance, \texttt{left-pad} pre-computes padding for strings below a certain length
and returns it directly. In contrast, \sys tends to produce straightforward
implementations without attempting similar optimizations.
\sys also omits domain-specific knowledge.
For example, in the \texttt{is} family,
\sys-generated code frequently fails to account for all edge cases and often misses JavaScript-specific patterns,
such as checking an object's string representation to decide if an object is circular, as in the \texttt{is-circular} components.
Reflecting current LLM capabilities,
\sys exhibits weak performance on components that involve complex string transformations or perform non-trivial math. 
In addition, \sys shows weak performance in higher-order functions (\eg \ttt{just-curry}, which implements function currying).
In these cases, the input generation module, while attempting to generate diverse inputs, results in input-output examples that use a variety of functions and thus are difficult for the program synthesis module to generalize from.
% Also, regenerating functions that themselves return functions is out-of-scope
% Also talk about math libraries with const tables
% and string libraries with complex regex

% How do we present each one of the libraries in the security evaluation set?

\vspace{-.1in}
\subsection{Performance}
\label{sec:performance}

For \textbf{Q3}, we evaluate the performance of \sys using \gptmodel
across the \totalLibs components in the evaluation set.
We measure how long it takes to regenerate a component
and the performance overhead of each regenerated component.
We report the wall-clock time difference between the start and 
end of the regeneration process,
split between the input-output pair generation and regeneration phases.
Additionally, to assess the performance overhead of the regenerated code 
compared to the original implementations,
we run the test suite of each component using both versions.
We repeat each performance experiment 100 times and report averages.

\sys takes an average of 1877s, taking a minimum of 676.54s on the \ttt{is-number} component and a maximum of 3858s on the \ttt{is-map} component to fully regenerate them,
with 826s on average spent on the generation of input/output pairs
and 1051s on program synthesis.
\sys is able to regenerate components with mostly negligible performance overhead, as it directly synthesis code in the target language.
The system incurs an average performance overhead of less than 0.1\% for the regenerated components, with a maximum overhead of 16.9\% on the \ttt{array-partition} component.
% with a minimum of 0\% and a maximum of 16.9\%.
On 41 instances, the regenerated component outperforms the original implementation.
% while on 13 instances, the performance overhead exceeds 100\%.
% On one instance (\ttt{decamelize}), the regenerated component timed out during test suite execution.
% Upon inspection, the regenerated component includes a bug that results in an infinite loop if the input string is empty.

% \begin{table}[h]
%     \caption{Native Third-party libraries in the evaluation set.}
%      \label{tbl:native-libs}
%   \centering
%   \begin{small}
%       \begin{tabular}{lrrc}
%       \hline
%           \toprule
%           \textbf{Name} & \textbf{LoC} & \textbf{RLoC} & \textbf{Tests} \\
%           \midrule
%           \ttt{String-Upper}~\cite{string-upper} & 150 & 120 & \greencmark \\
%           \ttt{Right-Trim}~\cite{right-trim} & 75 & 60 & \greencmark \\
%           \ttt{Left-Trim}~\cite{left-trim} & 80 & 65 & \greencmark \\
%           \ttt{Repeat-Text}~\cite{repeat-text} & 100 & 85 & \greencmark \\
%           \ttt{Character-Count}~\cite{character-count} & 120 & 100 & \greencmark \\
%           \ttt{Left-Right-Trim}~\cite{left-right-trim} & 90 & 70 & \greencmark \\
%           \ttt{Fill-Number}~\cite{fill-number} & 110 & 95 & \greencmark \\
%           \bottomrule
%       \end{tabular}
%   \end{small}
% \end{table}

\vspace{-.1in}
\subsection{\sysp vs. Harp (Program Synthesis)}
\label{sec:program-synthesis}

Harp~\cite{harp:ccs:2021} is a domain-specific program inference and regeneration system targeting components that perform string transformations.
Harp works with components whose behavior conforms to programs in a domain-specific language designed to support both (1) a domain-specific program inference algorithm and (2) a regeneration algorithm. 

For \textbf{Q5}, we include Harp's full benchmark set of components~\cite{harp:ccs:2021}
and compare \sys's performance and compatibility with Harp's domain-specific program inference and regeneration techniques. Applied to the 17 components from the Harp benchmark set~\cite{harp:ccs:2021}, 
  Harp's domain-specific synthesis engine and accompanying DSL
  outperforms \sys. \sys is able to correctly regenerate 
  5/17 components, while Harp regenerates 14/17.
We note that regular expression generation is a domain that is particularly challenging for \sys.

% Also relevant https://escholarship.org/content/qt8h27h8sb/qt8h27h8sb.pdf
% An approach to eliminate vulnerabilities in libraries is to use program synthesis~\cite{harp:ccs:2021}. These tools generate functionally equivalent versions of libraries from a set of input-output pairs. However, despite their effectiveness, they are often restricted in applicability across different domains and require a domain-specific language (DSL) to be defined for each new library domain.

In comparison with Harp, \sys 
can automatically regenerate components across a broader range of domains 
while maintaining high levels of correctness. 
For instance,
Harp is limited to regenerating components with string or stream processing components, 
such as \ttt{left-pad} and \ttt{trim}. 
\sys, on the other hand, 
is constrained only by the expressiveness
of the input-output pairs used to describe the component's functionality and the LLM regeneration capability. 

\subsection{\sysp vs. GPT-5 Mini (LLMs)}
\label{sec:eval-naive}

For \textbf{Q4}, we provide the source code of the same malicious components to \gptmodel with a prompt that instructs the model to generate
a functionally equivalent software component, with any malicious code removed~(\cref{sec:naive-prompt}).
We evaluate the security and functionality of the produced code in an identical manner as \sys's regeneration.

For components with simple functionality like \texttt{left-pad}, \texttt{trim}, and \texttt{string-compare},
\gptmodel generates a functionally equivalent version of the components, essentially by copying the legitimate code and discarding the malicious code.
In these cases, the generated code passes the developer-provided tests and is free of vulnerabilities, 
making this approach a viable alternative to \sys for these components.

% I want to discuss how this could be not so much that the LLM understands that code is malicious, but that it understands that the code is not part of the library's functionality, as it remembers this library or similar ones from its dataset.
% However, a key weakness of using pure LLMs to remove malicious behavior from source code
% is that this approach relies on the model's ability to \say{understand} what a vulnerability is.
% While this is often clear, such as when a library appends a pre-defined SSH key to the user's \texttt{.ssh/known\_hosts} file,
% sometimes this understanding requires more context, which might not be available.

In the case of the \texttt{pad} supply-chain attack~(\cref{sec:eval-pad}),
the malicious code snippet fetches unknown source code from a URL.
The LLM understands the call to \ttt{fetch} as benign, as it is a common operation and there 
is no context to suggest that the fetched code is malicious.
Similarly, if the vulnerability exists in a compiled module that is included using \texttt{require},
with only the component interface implemented in JavaScript,
as in the native addon vulnerabilities,
an LLM cannot ``see'' any
vulnerabilities in the compiled artifact and may simply copy the original code,
failing to provide a secure version of the component.
In both of these cases, \sys interacts with the component in a black-box manner,
generates a set of input-output pairs that adequately exercise the component's functionality,
and uses these pairs to regenerate a new version of the component.
For \ttt{pad}, the regenerated component includes only the padding functionality,
as fetching code from a URL is not representable by input-output pairs.

% Explain the process of turning a malicious snippet into one that looks safe
\heading{Prompt-based adversarial attack}
Prompt-based adversarial attacks~\cite{llmfool:2023} can also be used to trick
an LLM into classifying malicious code as benign and failing to remove it.
We employ this technique on \ttt{synthetic-left-pad} 
and \ttt{leetlog}.
The attack involves an adversary
(1) creating a component with an actively malicious snippet,
(2) providing the library source code to an LLM, 
(3) prompting the LLM to ``improve'' the code using ``best practices'', including error handling and safety checks while not modifying any of the malicious behavior, and
(4) repeating the process until the LLM no longer recognizes the malicious code as such.
In our setup, we use Mixtral~\cite{jiang2023mistral7b} to generate the adversarial examples,
providing it with the source code of the malicious components and a prompt 
that instructs the model to improve the code, adding error handling and safety checks.

In the case of \ttt{synthetic-left-pad}, Mixtral adds try-catch blocks around the malicious code that reads the contents of a file in the host file system and logs it to the console.
\gptmodel not only fails to recognize the vulnerabilities and malicious
snippets in these libraries, but
also does not remove them when prompted.
\sys synthesizes a secure, vulnerability-free version of the component by first generating a set of inputs to test the component's padding functionality.
Based on the resulting input-output pairs, it synthesizes a new version of the component that includes only the padding functionality.
Filesystem access is excluded, as it is not exercised during the input-output generation (the attack only activates if the \ttt{PRODUCTION} environment variable is set to true), and is unrepresentable in the input-output format.

\subsection{Ablation Study}
\label{sec:ablation}

We measure the effect of different features of the \sys implementation by eliminating the feature and measuring the
resulting ability of \sys to produce regenerations that pass both (1) the generated input-output pairs from the full \sys implementation and (2) the test suite for the component.
With full \sys, \sysCorrectLibs of \totalLibs components pass both.
Removing intermediate natural language algorithm generation (going directly from the input-output pairs to the final implementation), 62 of \totalLibs components pass both.
Removing library names, \sysCorrectLibs of \totalLibs pass both.
Removing input-output pairs, 27 of the \totalLibs pass both. These results highlight the importance of capturing legitimate component behavior via input-output pairs to enable successful component regeneration. 

% \begin{figure}[t]
%     \centering
%     \includegraphics[width=\columnwidth]{figs/microbenchmarks.pdf}
%     \caption{\textbf{Microbenchmark results.} 
%   Results on the impact of input-output pairs, an LLM-generated algorithm,
%   and the library name on regenerated component correctness. Base correctness is
%   under the default \sys configuration. The $y$-axis shows the percentage of
%   passed developer-provided tests.}
%     \label{fig:microbenchmarks}
%     \vspace{-1em}
% \end{figure}

\section{Related Work}
\label{sec:related-work}

Existing techniques for supply-chain security are not directly comparable to \sys, as they often focus on detecting security vulnerabilities, rather than removing them, or are limited to narrow domains or types of attacks.

\heading{LLMs for security}
Researchers have developed techniques that use LLMs to detect (and potentially eliminate) malicious code~\cite{alkaraki2024exploringllmsmalwaredetection, llmvulndetection2023, noever2023largelanguagemodelsfix, xu2024largelanguagemodelscyber, Yao_2024, akuthota2023vulnerability}. These techniques can fail, for example, when the malicious code has been constructed to successfully appear as legitimate or when an LLM has made the code appear to be legitimate~\sx{sec:eval-naive}. 
% In addition, LLMs on their own can not provide any form of guarantees on the correctness or security of the produced code.
\sys differs in that it uses input-output pairs to capture observable client-facing behavior, then uses these pairs to regenerate a completely new version. Instead of attempting to recognize and/or remove malicious code, \sys discards all of the original code, including any malicious code.

\heading{LLMs for test-case generation}
Test generation using LLMs
involves prompting an LLM to create unit tests for a software component \cite{empirical, github2023copilot, sapozhnikov2024testspark}, with some of this research generating tests to uncover software security vulnerabilities \cite{zhang2023does}.
\sys differs in that the generated input-output pairs are meant to capture the observable client-facing behavior of the component, rather than uncovering vulnerabilities or testing functionality---\sys's input-output format is constrained and includes only specific language primitives, avoiding the risk of encoding side-effectful malicious behaviors.

\heading{Component isolation}
Component isolation involves securing individual components through 
  sandboxing,
  interposition,
  wrapping, and access checks~\cite{mcc,
santos2014information,
ko2021securejs,
musch2019scriptprotect,
magazinius2014architectures,
agten2012jsand,
meyerovich2010conscript}.
\sys instead replaces the original component with a regenerated component that exhibits the same client-facing functionality. 
It does not use
sandboxing or wrapping to enforce security policies on the regenerated component
and thus has no runtime overhead during the production execution of a component.
\sys's sandboxing is used during inference when \sys interacts with a potentially malicious component.

\heading{Program synthesis}
Program synthesis, programming by example, and component-based synthesis
\cite{feser2015synthesizing,
gulwani2011automating,
raza2018disjunctive,
singh2016blinkfill,
polikarpova2016program,
alur2013syntax,
Handa_2020,
yaghmazadeh2018automated, 
jha2010oracle,
feng2017component,
shi2019frangel,
mandelin2005jungloid,
galenson2014codehint}
are methods for automatically generating programs,
either from high-level specifications,
example inputs and outputs,
or reusable components, respectively.
Techniques include constraint solving, refinement types, and interactive feedback.
\sys differs from these approaches in that 
(1) it uses an LLM to generate the input-output pairs (instead of working with a given set of input-output examples), 
  (2) the regeneration leverages an LLM instead of a domain-specific program synthesis algorithm, and 
  (3) programs produced during the regeneration phase do not have to conform to a domain-specific language (DSL).
Eliminating these constraints allows \sys's inference and regeneration to scale \emph{across} domains. Program synthesis, in contrast, targets specific domains (\eg string transforms, stream processing, array calculations, math operations) and requires both domain and programming language expertise to design and implement a domain-specific language for each targeted domain. %\sys's use of LLMs also significantly reduces engineering effort.

\heading{Program inference and regeneration}
Program inference and regeneration uses active learning applied to an existing implementation to infer a model of component (the model is a program in a domain specific language that captures computations in a targeted domain), then uses the model to regenerate a new implementation of the component~\cite{harp:ccs:2021,shen2019using}. The technique has been applied for software rejuvenation~\cite{shen2019using} and eliminating security vulnerabilities/malicious code~\cite{harp:ccs:2021}. This previous research targets specific domains as reflected in the need to design a domain specific language that (1) captures the target domain while (2) enabling the development of an active learning algorithm that can successfully infer computations within the domain. Our research, in contrast, uses LLMs to obtain input-output pairs that capture client-facing observable behavior and LLMs to regenerate the new implementation. This approach dramatically reduces the required programming language expertise and engineering effort and (by eliminating the need to focus on a specific domain) dramatically increases the scope of the system. 

\heading{Software de-bloating}
Software de-bloating~\cite{brown2024broad, azad2019less, Brown_2019, koishybayev2020mininode} is a
technique that removes unused or unnecessary code from software components. This
is typically achieved by analyzing their execution under controlled
environments or performing reachability analysis to identify and eliminate code that is not
invoked or required for the observed functionality.
\sys differs in that (1) it does not remove specific code segments, but rather
regenerates the entire component from scratch and (2) \sys eliminates malicious code (including code that may execute during inference) 
as long as this code does not affect the observable behavior of the target component (as opposed to eliminating unused or unnecessary code). 

\heading{Vulnerability detection}
Static and dynamic analysis (and their combination)~
\cite{calzavara2015fine, 
maffeis2009language,
fass2019jstap,
xu2013jstill,
Jamrozik2016,
Koved2002,
lya:fse:2021,
stokes2019scriptnet,
schutt2012early,
mao2018detecting,
wang2018combined,
dewald2010adsandbox,
hu2018jsforce,
phung2024jsmbox,
mirccs2021}
attempt to detect vulnerabilities present in a given software artifact or infer information about the behavior of a program to enforce it at runtime.
Instead of attempting to infer malicious or benign behavior, \sys completely regenerates a component into a vulnerability-free version with equivalent behavior in regards to its client-observable functionality.

\vspace{-.2in}
\section{Discussion \& Limitations}
\vspace{-.1in}
\label{sec:limitations}

% \sys comes with a number of limitations.
%that open up exciting possibilities for future work.

% \heading{Common Failure Modes}
% Cases where \sys fails to regenerate the given library can be summarized
% into the following failure modes:
% (i) The generated input-output pairs, while correct, do not adequately represent the original code,
% (ii) the source code cannot be regenerated correctly, even when the generated input-output pairs achieve high code coverage, and
% (iii) the library's behavior is difficult or impossible to represent in \sys's input-output pair format.

\heading{Operation Granularity}
The current version of \sys operates at the function level within individual components, regenerating pure functions that they expose.
% These functions do not have side-effects as part of their intended behavior (\ie file-system or network access).
\sys is able to detect if a component falls outside this scope during regeneration, at which point it will halt
and produce an error message.
The component's client-visible behavior should be representable by \sys's intermediate input-output pair format.
\sys focuses on regenerating components that perform arbitrary computations and (given current LLM capabilities) are typically around or below 150 lines of code.
% These libraries are widely used, as they are often dependencies to popular user-facing software.
The current \sys prototype has been implemented to regenerate components written in JavaScript, Python, Ruby, or wrapped as a native NodeJS module and implemented in any language.
 
\heading{Attacks on return values}
Consistent with its threat model~\sx{sec:threat-model}, \sys does not protect against malicious code that changes the results that components return to the client (unless the malicious code is activated only under specific conditions that do not hold during \sys inference and regeneration).
In this scenario, the malicious behavior will be modeled in the generated input-output pairs and regenerated in the new component. 

% malicious outputs. \sys does not aim to defend against a malicious library that returns incorrect outputs when certain conditions are met---malicious input/output behavior will be modeled as part of \sys's IO pairs and eventually be regenerated.
% an object, whose methods perform a malicious attack, \sys might insert this unwanted functionality in the regenerated library.

\heading{Concealing benign behaviors}
% \sys is not immune to adversarial attacks on the source code.
If a component's source is modified to conceal or affect its legitimate behavior, then \sys's input-output pairs may not fully capture the intended behavior of the component---resulting in incorrect code generation that fails developer tests.
In this case, \sys will issue an error, indicating inability to regenerate the component.
%Such overt attacks can be detected and mitigated using conventional analysis and transformation approaches~\sx{sec:related-work}.

\heading{Extending \sys to more languages}
% The \sys prototype is currently implemented to regenerate JavaScript code.
Extending \sys to target a new language requires adding
  (1) new, slightly modified, prompts to instruct the LLM back-end to generate code in the target language,
  (2) an appropriate code-coverage subsystem analyzing and reporting component coverage for the target language,
  (3) serialization and deserialization support for the input-output pairs in the target language.
In our experience, all three can be implemented in about an afternoon for popular languages that offer code-coverage and serialization functionality either built-in or via third-party components.

% Such attacks might involve techniques such as renaming of variables,
% adding dead code, or adding irrelevant code snippets.

\section{Conclusion}
\label{sec:conclusion}
Supply-chain attacks on software repositories are an escalating concern in today's digital landscape. 
This work presents \sys, a novel system
that automatically learns and regenerates potentially malicious components into safe versions with no malicious code.
A critical aspect of \sys is its use of input-output pairs to 
model component behavior, a design choice that severs any 
connection between the original, potentially malicious, code
and the regenerated code.  Our results highlight the 
effectiveness of \sys in learning, modeling, and regenerating
fully functional components from a range of domains. As such,
\sys provides a promising new direction for eliminating software supply chain vulnerabilities.

% \sys instruments multiple LLM instances, reinforced with guardrails to ensure correctness and completeness.\sys achieves high levels of correctness and performance across domains.

%This work is a crucial step towards regenerating safe versions of entire software ecosystems, \ie ones that minimize the risks of supply-chain incidents in a language- and domain-agnostic fashion.

\subsection*{Acknowledgements}
% \begin{acks}
  This work is supported by the National Science Foundation (NSF) through award
  CNS-2247687, the Defense Advanced Research Projects Agency (DARPA), through
  contract HR001120C0191, a Brown Computer Science Faculty Innovators Fund, and
  a Google ML-and-Systems Junior Faculty Award.
  Any opinions, findings, conclusions, or recommendations expressed herein are
  those of the authors and do not necessarily reflect the views of NSF or
  DARPA.
% \end{acks}

\bibliographystyle{plain}

%% Bibliography
{\small
\bibliography{./bib}
}

% \onecolumn
\appendix

\section{Prompts used by \sys}
\label{appendix:prompts}

This section presents the prompts used internally by \sys when targeting JavaScript components.

\begin{prompt-box}{Input set generation}
Given a component, generate an input test suite that will test thoroughly the
component's behavior. These input/output pairs will then be used to regenerate the module. Make
sure to include all edge cases and key behaviors.

1) Code Understanding: Explain the code's purpose and functionality. Identify key behaviors that require testing.

2) Inputs and Outputs: Brainstorm concise tests that are not repetitive for input/output
correctness, including key behaviors as well as edge cases. Make sure to test default arguments
and optional parameters. Try and find cases where the function returns a function, and test
that function as well.

3) Errors: Incorporate test cases triggering errors or exceptions for inputs where the provided
module throws errors.

4) Explore: Include tests that are slight variations of the ones in the test suite. Modify the
inputs slightly to test the function's behavior with different inputs.

5) Format: Output one object per line with language primitives as values. If the expected output is a
function, also provide a set of arguments to call that function with. Wrap all these argument
objects in an array.

Use this structure:

\begin{Verbatim}
Code: {Input component source code}
\end{Verbatim}
\end{prompt-box}

\begin{prompt-box}{Input-output pairs to natural language algorithm}
Given this test suite of a function, design an algorithm that describes the function.

1) Understand the Test Suite: Familiarize yourself with the requirements, inputs, outputs, and
any implicit criteria in the test cases.

2) Analyze the Problem: Define the problem based on the test suite, understanding applicable
concepts and complexity constraints.

3) Design the Algorithm: Develop a step-by-step approach, selecting suitable algorithms and
ensuring coverage of test case scenarios.

4) Handle Edge Cases: Identify and address any edge cases not explicitly covered in the test suite.

\begin{Verbatim}
I/O Pairs: {Input-output pairs}
\end{Verbatim}
\end{prompt-box}

\begin{prompt-box}{Component regeneration}
Generate a function given a set of input-output examples.

1) Understanding Test Specifications: Before writing code, thoroughly understand what each test
in the suite is checking. Know the input and output requirements, and what constitutes a pass
or fail.

2) Functional Correctness: Ensure your code meets the functional requirements outlined in the
tests. It should correctly handle all specified cases, including edge cases.

3) Code Quality: Write clean, readable, and well-structured code. Use descriptive variable
names, avoid hard-coding, and follow best practices. 

4) Context: You might need to understand the context of the function, such as the purpose of
the component it belongs to, to write the function correctly.

5) Refactoring: After your code passes the tests, look for opportunities to refactor. Simplify
complex parts, remove duplicated code, and improve the overall structure.

\begin{Verbatim}
LIbrary Name: {Input library name}
I/O Pairs: {Generated examples}
Algorithm: {Input algorithm}
\end{Verbatim}
\end{prompt-box}

\section{Naive LLM Approach Prompt}
\label{sec:naive-prompt}

This section presents the prompt used to evaluate a naive approach to removing malicious code from a component using a state-of-the-art LLM.

\begin{prompt-box}{Naive approach}
Given the code for a potentially malicious function, remove any security vulnerabilities without compromising the functionality of the code.

1) Identify any potentially malicious snippets or vulnerabilities in the code.

2) Modify the code to remove the security vulnerability without compromising the functionality of the code.

3) Return the code with the security vulnerabilities removed.
\begin{Verbatim}
{Input source code}
\end{Verbatim}
\end{prompt-box}

\end{document}

%% file: table.tex
\begin{tabular*}{\textwidth}{lllrrrlrrrlllr}
  \toprule
\multicolumn{3}{c}{Library}   & \multicolumn{2}{c}{Popularity} & \multicolumn{4}{c}{I/O Pair Generation} & \multicolumn{4}{c}{Program Synthesis} & \\
Name & $\mathcal{D}$ & $\mathcal{L}$ & Weekly & Deps & $t_{gen}$ (s) & $R_{i}$ & P & Covg & $t_{synth}$ (s) & $R_{s}$ & Correct$_i$ & Correct$_d$ & Perf/nce \\
  \midrule
 \texttt{arr-diff} & A & JS & 18.31M & 9.9K & 266 & 1 & 24 & 100\% & 957 & 1 & \textcolor{black}{24/24} & \textcolor{black}{8/8} & < 0.1\% \\
 \texttt{is-number} & I & JS & 108.24M & 0.16M & 302 & 1 & 40 & 100\% & 374 & 1 & \textcolor{black}{40/40} & \textcolor{black}{111/111} & 2.4\% \\
 \texttt{has-proto} & H & JS & 48.06M & 2.8K & 412 & 1 & 29 & 100\% & 367 & 1 & \textcolor{black}{29/29} & \textcolor{black}{2/2} & < 0.1\% \\
 \texttt{primality} & M & PY & 13 & -- & 314 & 512 & 3 & 100\% & 2338 & 20 & \textcolor{black}{20/20} & \textcolor{black}{9/9} & 3\% \\
 \texttt{split-on-first} & S & JS & 9.35M & 0.79M & 499 & 1 & 35 & 100\% & 1273 & 1 & \textcolor{black}{35/35} & \textcolor{black}{7/7} & 3.2\% \\
 \texttt{just-filter-object} & O & JS & 9.9K & 0.35M & 649 & 1 & 15 & 100\% & 711 & 2 & \textcolor{black}{15/15} & \textcolor{black}{3/3} & 1.1\% \\
 \texttt{is-odd} & M & JS & 542.3K & 1.8K & 1205 & 1 & 32 & 100\% & 985 & 1 & \textcolor{black}{32/32} & \textcolor{black}{3/3} & 3.6\% \\
 \texttt{concat-map} & C & JS & 80.41M & 43.3K & 1264 & 3 & 34 & 100\% & 1585 & 2 & \textcolor{black}{34/34} & \textcolor{black}{5/5} & < 0.1\% \\
 \texttt{array-ify} & A & JS & 8.88M & 158 & 303 & 1 & 25 & 100\% & 987 & 2 & \textcolor{black}{25/25} & \textcolor{black}{6/6} & 3.4\% \\
  \ttt{fast\_blank} & S & RB & 106,001 & 20 & 310 & 1 & 12 & 100\% & 432 & 1 & \textcolor{black}{12/12} & \textcolor{black}{4/4} & < 0.1\% \\
 \texttt{is-object} & I & JS & 4.01M & 0.28M & 359 & 1 & 33 & 100\% & 1309 & 2 & \textcolor{black}{33/33} & \textcolor{black}{9/9} & < 0.1\% \\
 \texttt{replace-ext} & S & JS & 7.96M & 9.4K & 545 & 1 & 36 & 100\% & 832 & 2 & \textcolor{black}{36/36} & \textcolor{black}{8/8} & < 0.1\% \\
 \texttt{just-pick} & O & JS & 30.9K & 9.1K & 650 & 1 & 20 & 100\% & 917 & 2 & \textcolor{black}{20/20} & \textcolor{black}{6/6} & < 0.1\% \\
 \texttt{is-even} & M & JS & 161.6K & 38.2K & 721 & 1 & 41 & 94\% & 1237 & 1 & \textcolor{black}{41/41} & \textcolor{black}{4/4} & 2.9\% \\
  \texttt{character-count} & S & C++ & 2 & -- & 1254 & 1 & 9 & 100\% & 2338 & 1 & \textcolor{black}{9/9} & \textcolor{black}{5/5} & 4.3\% \\
\midrule
\multicolumn{14}{l}{{\ldots 135 more components}} \\
\midrule
Min &  &  & 1 & 4 & 265 & 1 & 0 & 4\% & 171 & 1 & 0\% & 0\% & < 0.1\% \\
Max &  &  & 108.24M & 1.47M & 2225 & 3 & 100 & 100\% & 2338 & 3 & 100\% & 100\% & 16.9\% \\
Avg &  &  & 9.56M & 0.19M & 826 & 1 & 32 & 93\% & 1051 & 1 & 94\% & 93\% & < 0.1\% \\
  \bottomrule
\end{tabular*}

%% file: draft.bbl
\begin{thebibliography}{10}

\bibitem{strace}
strace - trace system calls and signals.
\newblock \url{https://man7.org/linux/man-pages/man1/strace.1.html}, 2025.
\newblock Accessed: 2025-10-22.

\bibitem{agten2012jsand}
Pieter Agten, Steven Van~Acker, Yoran Brondsema, Phu~H. Phung, Lieven Desmet,
  and Frank Piessens.
\newblock {JSand: Complete Client-side Sandboxing of Third-party JavaScript
  Without Browser Modifications}.
\newblock In {\em Proceedings of the 28th Annual Computer Security Applications
  Conference}, ACSAC '12, pages 1--10, New York, NY, USA, 2012. ACM.

\bibitem{akuthota2023vulnerability}
Vishwanath Akuthota, Raghunandan Kasula, Sabiha~Tasnim Sumona, Masud Mohiuddin,
  Md~Tanzim Reza, and Md~Mizanur Rahman.
\newblock {Vulnerability Detection and Monitoring Using LLM}.
\newblock In {\em 2023 IEEE 9th International Women in Engineering (WIE)
  Conference on Electrical and Computer Engineering (WIECON-ECE)}, pages
  309--314. IEEE, 2023.

\bibitem{alkaraki2024exploringllmsmalwaredetection}
Jamal Al-Karaki, Muhammad Al-Zafar Khan, and Marwan Omar.
\newblock {Exploring LLMs for Malware Detection: Review, Framework Design, and
  Countermeasure Approaches}, 2024.

\bibitem{alur2013syntax}
Rajeev Alur, Rastislav Bodik, Garvit Juniwal, Milo~MK Martin, Mukund
  Raghothaman, Sanjit~A Seshia, Rishabh Singh, Armando Solar-Lezama, Emina
  Torlak, and Abhishek Udupa.
\newblock {Syntax-guided synthesis}.
\newblock In {\em 2013 Formal Methods in Computer-Aided Design}, pages 1--8.
  IEEE, 2013.

\bibitem{ev:eurosec:2022}
Iosif Arvanitis, Grigoris Ntousakis, Sotiris Ioannidis, and Nikos Vasilakis.
\newblock {A Systematic Analysis of the Event-Stream Incident}.
\newblock In {\em 15th European Workshop on Systems Security}, EuroSec '22, New
  York, NY, USA, 2022. Association for Computing Machinery.

\bibitem{azad2019less}
Babak~Amin Azad, Pierre Laperdrix, and Nick Nikiforakis.
\newblock {Less is more: quantifying the security benefits of debloating web
  applications}.
\newblock In {\em 28th $\{$USENIX$\}$ Security Symposium ($\{$USENIX$\}$
  Security 19)}, pages 1697--1714, 2019.

\bibitem{copay}
{BITPAY INC}.
\newblock {Copay}.
\newblock \url{https://github.com/bitpay/copay/}, 2015.
\newblock Accessed: 2021-09-09.

\bibitem{brown2024broad}
Michael~D. Brown, Adam Meily, Brian Fairservice, Akshay Sood, Jonathan Dorn,
  Eric Kilmer, and Ronald Eytchison.
\newblock {A Broad Comparative Evaluation of Software Debloating Tools}, 2024.

\bibitem{Brown_2019}
Michael~D. Brown and Santosh Pande.
\newblock {CARVE: Practical Security-Focused Software Debloating Using Simple
  Feature Set Mappings}.
\newblock In {\em Proceedings of the 3rd ACM Workshop on Forming an Ecosystem
  Around Software Transformation}, CCS ’19. ACM, November 2019.

\bibitem{calzavara2015fine}
Stefano Calzavara, Michele Bugliesi, Silvia Crafa, and Enrico Steffinlongo.
\newblock {Fine-grained detection of privilege escalation attacks on browser
  extensions}.
\newblock In {\em Programming Languages and Systems: 24th European Symposium on
  Programming, ESOP 2015, Held as Part of the European Joint Conferences on
  Theory and Practice of Software, ETAPS 2015, London, UK, April 11-18, 2015,
  Proceedings 24}, pages 510--534. Springer, 2015.

\bibitem{alr}
Jos\'{e}~P. Cambronero, Thurston H.~Y. Dang, Nikos Vasilakis, Jiasi Shen, Jerry
  Wu, and Martin~C. Rinard.
\newblock {Active Learning for Software Engineering}.
\newblock In {\em Proceedings of the 2019 ACM SIGPLAN International Symposium
  on New Ideas, New Paradigms, and Reflections on Programming and Software},
  Onward! 2019, page 62–78, New York, NY, USA, 2019. Association for
  Computing Machinery.

\bibitem{Aspire_2015}
Kevin Chen, Warren He, Devdatta Akhawe, Vijay D{ \textquoteright}Silva, Prateek
  Mittal, and Dawn Song.
\newblock {{ASPIRE}: Iterative Specification Synthesis for Security}.
\newblock In {\em 15th Workshop on Hot Topics in Operating Systems (HotOS XV)},
  Kartause Ittingen, Switzerland, May 2015. USENIX Association.

\bibitem{chen2018synthesizing}
Xinyun Chen, Chang Liu, and Dawn Song.
\newblock {Towards Synthesizing Complex Programs from Input-Output Examples},
  2018.

\bibitem{cloudflare2023}
Cloudflare.
\newblock {SolarWinds Orion Compromise Trend Data}, 2023.
\newblock Accessed: 2024-10-30.

\bibitem{Quartz2016}
Keith Collins.
\newblock {How one programmer broke the internet by deleting a tiny piece of
  code}, 2016.

\bibitem{dewald2010adsandbox}
Andreas Dewald, Thorsten Holz, and Felix~C Freiling.
\newblock {ADSandbox: Sandboxing JavaScript to fight malicious websites}.
\newblock In {\em proceedings of the 2010 ACM Symposium on Applied Computing},
  pages 1859--1864, 2010.

\bibitem{cybersecuritydive_3cx_attack_2024}
Cybersecurity Dive.
\newblock {Supply Chain Attack on 3CX Desktop Software Impacts Thousands},
  2024.
\newblock Accessed: 2024-11-05.

\bibitem{eventStreamNPM}
{Dominic Tarr}.
\newblock {Event-Stream}.
\newblock \url{https://www.npmjs.com/package/event-stream}, 2011.
\newblock Accessed: 2021-09-09.

\bibitem{fass2019jstap}
Aurore Fass, Michael Backes, and Ben Stock.
\newblock {Jstap: a static pre-filter for malicious javascript detection}.
\newblock In {\em Proceedings of the 35th Annual Computer Security Applications
  Conference}, pages 257--269, 2019.

\bibitem{feng2024towards}
Guhao Feng, Bohang Zhang, Yuntian Gu, Haotian Ye, Di~He, and Liwei Wang.
\newblock {Towards revealing the mystery behind chain of thought: a theoretical
  perspective}.
\newblock {\em {Advances in Neural Information Processing Systems}}, 36, 2024.

\bibitem{feng2017component}
Yu~Feng, Ruben Martins, Yuepeng Wang, Isil Dillig, and Thomas~W Reps.
\newblock {Component-based synthesis for complex APIs}.
\newblock In {\em Proceedings of the 44th ACM SIGPLAN Symposium on Principles
  of Programming Languages}, pages 599--612, 2017.

\bibitem{feser2015synthesizing}
John~K Feser, Swarat Chaudhuri, and Isil Dillig.
\newblock {Synthesizing data structure transformations from input-output
  examples}.
\newblock In {\em ACM SIGPLAN Notices}, volume~50, pages 229--239. ACM, 2015.

\bibitem{Foy_2019}
David Foy.
\newblock {The Frightening State of Security Around npm Package Management}.
\newblock \url{
  https://naildrivin5.com/blog/2019/07/10/the-frightening-state-security-around-npm-package-management.html
  }, July 2019.

\bibitem{galenson2014codehint}
Joel Galenson, Philip Reames, Rastislav Bodik, Bj{\"o}rn Hartmann, and Koushik
  Sen.
\newblock {Codehint: Dynamic and interactive synthesis of code snippets}.
\newblock In {\em Proceedings of the 36th International Conference on Software
  Engineering}, pages 653--663, 2014.

\bibitem{github2023copilot}
{GitHub}.
\newblock {About GitHub Copilot Chat in your IDE}, 2023.
\newblock Accessed: 2024-06-22.

\bibitem{github_advisory_GHSA-gfm8-g3vm-53jh}
{GitHub}.
\newblock {GHSA-gfm8-g3vm-53jh: Cross-Site Scripting in SimpleMDE}.
\newblock \url{https://github.com/advisories/GHSA-gfm8-g3vm-53jh}, 2023.
\newblock Accessed: 2023-10-31.

\bibitem{wired_barium_supply_chain}
Andy Greenberg.
\newblock {A Mysterious Hacker Group Is On a Supply Chain Hijacking Spree}.
\newblock {\em {Wired}}, May 2019.
\newblock Accessed: 2023-10-20.

\bibitem{gulwani2011automating}
Sumit Gulwani.
\newblock {Automating string processing in spreadsheets using input-output
  examples}.
\newblock In {\em ACM Sigplan Notices}, volume~46, pages 317--330. ACM, 2011.

\bibitem{Handa_2020}
Shivam Handa and Martin~C. Rinard.
\newblock {Inductive program synthesis over noisy data}.
\newblock {\em {Proceedings of the 28th ACM Joint Meeting on European Software
  Engineering Conference and Symposium on the Foundations of Software
  Engineering}}, Nov 2020.

\bibitem{hu2018jsforce}
Xunchao Hu, Yao Cheng, Yue Duan, Andrew Henderson, and Heng Yin.
\newblock {Jsforce: A forced execution engine for malicious javascript
  detection}.
\newblock In {\em Security and Privacy in Communication Networks: 13th
  International Conference, SecureComm 2017, Niagara Falls, ON, Canada, October
  22--25, 2017, Proceedings 13}, pages 704--720. Springer, 2018.

\bibitem{hugeglass2024flatmapstream}
hugeglass.
\newblock {flatmap-stream}.
\newblock \url{https://github.com/hugeglass/flatmap-stream}, 2024.
\newblock Accessed: 2024-06-19.

\bibitem{Jamrozik2016}
Konrad Jamrozik, Philipp von Styp{-}Rekowsky, and Andreas Zeller.
\newblock {Mining sandboxes}.
\newblock In Laura~K. Dillon, Willem Visser, and Laurie~A. Williams, editors,
  {\em Proceedings of the 38th International Conference on Software
  Engineering, {ICSE} 2016, Austin, TX, USA, May 14-22, 2016}, pages 37--48.
  {ACM}, 2016.

\bibitem{jha2010oracle}
Susmit Jha, Sumit Gulwani, Sanjit~A Seshia, and Ashish Tiwari.
\newblock {Oracle-guided component-based program synthesis}.
\newblock In {\em Proceedings of the 32nd ACM/IEEE International Conference on
  Software Engineering-Volume 1}, pages 215--224, 2010.

\bibitem{jiang2023mistral7b}
Albert~Q. Jiang, Alexandre Sablayrolles, Arthur Mensch, Chris Bamford,
  Devendra~Singh Chaplot, Diego de~las Casas, Florian Bressand, Gianna Lengyel,
  Guillaume Lample, Lucile Saulnier, Lélio~Renard Lavaud, Marie-Anne Lachaux,
  Pierre Stock, Teven~Le Scao, Thibaut Lavril, Thomas Wang, Timothée Lacroix,
  and William~El Sayed.
\newblock {Mistral 7B}, 2023.

\bibitem{khattab2023dspy}
Omar Khattab, Arnav Singhvi, Paridhi Maheshwari, Zhiyuan Zhang, Keshav
  Santhanam, Sri Vardhamanan, Saiful Haq, Ashutosh Sharma, Thomas~T. Joshi,
  Hanna Moazam, Heather Miller, Matei Zaharia, and Christopher Potts.
\newblock {DSPy: Compiling Declarative Language Model Calls into Self-Improving
  Pipelines}.
\newblock {\em {arXiv preprint arXiv:2310.03714}}, 2023.

\bibitem{khot2023decomposed}
Tushar Khot, Harsh Trivedi, Matthew Finlayson, Yao Fu, Kyle Richardson, Peter
  Clark, and Ashish Sabharwal.
\newblock {Decomposed Prompting: A Modular Approach for Solving Complex Tasks}.
\newblock In {\em The Eleventh International Conference on Learning
  Representations}, 2023.

\bibitem{ko2021securejs}
Yoonseok Ko, Tamara Rezk, and Manuel Serrano.
\newblock {SecureJS Compiler: Portable Memory Isolation in JavaScript}.
\newblock In {\em SAC 2021-The 36th ACM/SIGAPP Symposium On Applied Computing},
  2021.

\bibitem{koishybayev2020mininode}
Igibek Koishybayev and Alexandros Kapravelos.
\newblock {Mininode: Reducing the Attack Surface of Node.js Applications}.
\newblock In {\em 23rd International Symposium on Research in Attacks,
  Intrusions and Defenses ($\{$RAID$\}$ 2020)}, 2020.

\bibitem{Koved2002}
Larry Koved, Marco Pistoia, and Aaron Kershenbaum.
\newblock {Access rights analysis for Java}.
\newblock In Mamdouh Ibrahim and Satoshi Matsuoka, editors, {\em Proceedings of
  the 2002 {ACM} {SIGPLAN} Conference on Object-Oriented Programming Systems,
  Languages and Applications, {OOPSLA} 2002, Seattle, Washington, USA, November
  4-8, 2002}, pages 359--372. {ACM}, 2002.

\bibitem{li2023structured}
Jia Li, Ge~Li, Yongmin Li, and Zhi Jin.
\newblock {Structured chain-of-thought prompting for code generation}.
\newblock {\em {ACM Transactions on Software Engineering and Methodology}},
  2023.

\bibitem{maffeis2009language}
Sergio Maffeis and Ankur Taly.
\newblock {Language-based isolation of untrusted Javascript}.
\newblock In {\em 2009 22nd IEEE Computer Security Foundations Symposium},
  pages 77--91. IEEE, 2009.

\bibitem{magazinius2014architectures}
Jonas Magazinius, Daniel Hedin, and Andrei Sabelfeld.
\newblock {Architectures for inlining security monitors in web applications}.
\newblock In {\em International Symposium on Engineering Secure Software and
  Systems}, pages 141--160. Springer, 2014.

\bibitem{mandelin2005jungloid}
David Mandelin, Lin Xu, Rastislav Bod{\'\i}k, and Doug Kimelman.
\newblock {Jungloid mining: helping to navigate the API jungle}.
\newblock {\em {ACM Sigplan Notices}}, 40(6):48--61, 2005.

\bibitem{mao2018detecting}
Jian Mao, Jingdong Bian, Guangdong Bai, Ruilong Wang, Yue Chen, Yinhao Xiao,
  and Zhenkai Liang.
\newblock {Detecting malicious behaviors in javascript applications}.
\newblock {\em {Ieee Access}}, 6:12284--12294, 2018.

\bibitem{meyerovich2010conscript}
Leo~A Meyerovich and Benjamin Livshits.
\newblock {ConScript: Specifying and enforcing fine-grained security policies
  for Javascript in the browser}.
\newblock In {\em 2010 IEEE Symposium on Security and Privacy}, pages 481--496.
  IEEE, 2010.

\bibitem{musch2019scriptprotect}
Marius Musch, Marius Steffens, Sebastian Roth, Ben Stock, and Martin Johns.
\newblock {ScriptProtect: mitigating unsafe third-party javascript practices}.
\newblock In {\em Proceedings of the 2019 ACM Asia Conference on Computer and
  Communications Security}, pages 391--402, 2019.

\bibitem{cve:strong-password}
{National Vulnerability Database (NVD)}.
\newblock {CVE-2019-13354 Detail: The strong\_password gem 0.0.7 for Ruby...
  included a code-execution backdoor inserted by a third party.}
\newblock \url{https://nvd.nist.gov/vuln/detail/CVE-2019-13354}, July 2019.
\newblock Accessed: 2025-12-06.

\bibitem{noever2023largelanguagemodelsfix}
David Noever.
\newblock {Can Large Language Models Find And Fix Vulnerable Software?}, 2023.

\bibitem{es2}
{npm, Inc.}
\newblock {Details about the event-stream incident}.
\newblock \url{
  https://blog.npmjs.org/post/180565383195/details-about-the-event-stream-incident
  }, 2018.
\newblock Accessed: 2018-12-18.

\bibitem{nye2021work}
Maxwell Nye, Anders~Johan Andreassen, Guy Gur-Ari, Henryk Michalewski, Jacob
  Austin, David Bieber, David Dohan, Aitor Lewkowycz, Maarten Bosma, David
  Luan, Charles Sutton, and Augustus Odena.
\newblock {Show Your Work: Scratchpads for Intermediate Computation with
  Language Models}, 2021.

\bibitem{ohm2020backstabber}
Marc Ohm, Henrik Plate, Arnold Sykosch, and Michael Meier.
\newblock {Backstabber's Knife Collection: A Review of Open Source Software
  Supply Chain Attacks}.
\newblock In {\em International Conference on Detection of Intrusions and
  Malware , and Vulnerability Assessment}. Springer, 2020.

\bibitem{openai2024gpt4ocard}
OpenAI.
\newblock {GPT-4o} system card, 2024.

\bibitem{gpt5systemcard}
{OpenAI}.
\newblock {GPT-5} system card, 2025.

\bibitem{park2024grammaraligneddecoding}
Kanghee Park, Jiayu Wang, Taylor Berg-Kirkpatrick, Nadia Polikarpova, and Loris
  D'Antoni.
\newblock {Grammar-Aligned Decoding}, 2024.

\bibitem{phung2024jsmbox}
Phu~H Phung, Allen Varghese, Bojue Wang, Yu~Zhao, and Chong Yu.
\newblock {JSMBox—A Runtime Monitoring Framework for Analyzing and
  Classifying Malicious JavaScript}.
\newblock In {\em International Conference on Software Engineering and Data
  Engineering}, pages 100--122. Springer, 2024.

\bibitem{polikarpova2016program}
Nadia Polikarpova, Ivan Kuraj, and Armando Solar-Lezama.
\newblock {Program synthesis from polymorphic refinement types}.
\newblock {\em {ACM SIGPLAN Notices}}, 51(6):522--538, 2016.

\bibitem{llmvulndetection2023}
Moumita~Das Purba, Arpita Ghosh, Benjamin~J. Radford, and Bill Chu.
\newblock {Software Vulnerability Detection using Large Language Models}.
\newblock In {\em 2023 IEEE 34th International Symposium on Software
  Reliability Engineering Workshops (ISSREW)}, pages 112--119, 2023.

\bibitem{raza2018disjunctive}
Mohammad Raza and Sumit Gulwani.
\newblock {Disjunctive Program Synthesis: A Robust Approach to Programming by
  Example}.
\newblock In {\em Thirty-Second AAAI Conference on Artificial Intelligence},
  2018.

\bibitem{promptsurvey:arxiv:2024}
Pranab Sahoo, Ayush~Kumar Singh, Sriparna Saha, Vinija Jain, Samrat Mondal, and
  Aman Chadha.
\newblock {A Systematic Survey of Prompt Engineering in Large Language Models:
  Techniques and Applications}.
\newblock Feb 2024.

\bibitem{sansec_polyfill_2023}
Sansec.
\newblock {Polyfill Supply Chain Attack}, 2023.
\newblock Accessed: 2024-10-30.

\bibitem{santos2014information}
Jos{\'e}~Fragoso Santos and Tamara Rezk.
\newblock {An information flow monitor-inlining compiler for securing a core of
  javascript}.
\newblock In {\em IFIP International Information Security Conference}, pages
  278--292. Springer, 2014.

\bibitem{saparov2022language}
Abulhair Saparov and He~He.
\newblock {Language models are greedy reasoners: A systematic formal analysis
  of chain-of-thought}.
\newblock {\em {arXiv preprint arXiv:2210.01240}}, 2022.

\bibitem{sapozhnikov2024testspark}
Arkadii Sapozhnikov, Mitchell Olsthoorn, Annibale Panichella, Vladimir
  Kovalenko, and Pouria Derakhshanfar.
\newblock {TestSpark: IntelliJ IDEA's Ultimate Test Generation Companion},
  2024.

\bibitem{schutt2012early}
Kristof Sch{\"u}tt, Marius Kloft, Alexander Bikadorov, and Konrad Rieck.
\newblock {Early detection of malicious behavior in javascript code}.
\newblock In {\em Proceedings of the 5th ACM Workshop on Security and
  Artificial Intelligence}, pages 15--24, 2012.

\bibitem{empirical}
Max Schäfer, Sarah Nadi, Aryaz Eghbali, and Frank Tip.
\newblock {An Empirical Evaluation of Using Large Language Models for Automated
  Unit Test Generation}, 2023.

\bibitem{mcc}
R.~Sekar, V.N. Venkatakrishnan, Samik Basu, Sandeep Bhatkar, and Daniel~C.
  DuVarney.
\newblock {Model-Carrying Code: A Practical Approach for Safe Execution of
  Untrusted Applications}.
\newblock In {\em Proceedings of the Twenty-Fourth ACM Symposium on Operating
  Systems Principles}, SOSP '03, pages 15--28, New York, NY, USA, 2003.
  Association for Computing Machinery.

\bibitem{Sharma2022ctx}
Ax~Sharma.
\newblock {PyPI Package 'ctx' and PHP Library 'phpass' Compromised to Steal
  Environment Variables}.
\newblock
  \url{https://www.sonatype.com/blog/pypi-package-ctx-compromised-are-you-at-risk},
  May 2022.
\newblock Accessed: 2025-12-06.

\bibitem{shen2019using}
Jiasi Shen and Martin~C. Rinard.
\newblock {Using Active Learning to Synthesize Models of Applications That
  Access Databases}.
\newblock In {\em Proceedings of the 40th ACM SIGPLAN Conference on Programming
  Language Design and Implementation}, PLDI 2019, pages 269--285, New York, NY,
  USA, 2019. ACM.

\bibitem{shi2019frangel}
Kensen Shi, Jacob Steinhardt, and Percy Liang.
\newblock {FrAngel: component-based synthesis with control structures}.
\newblock {\em {Proceedings of the ACM on Programming Languages}},
  3(POPL):1--29, 2019.

\bibitem{singh2016blinkfill}
Rishabh Singh.
\newblock {Blinkfill: Semi-supervised programming by example for syntactic
  string transformations}.
\newblock {\em {Proceedings of the VLDB Endowment}}, 9(10):816--827, 2016.

\bibitem{snyk-eventstream-2018}
Snyk.
\newblock {SNYK-JS-EVENTSTREAM-72638: Malicious Package in event-stream},
  November 2018.
\newblock Accessed: 2024-06-09.

\bibitem{snyk-flatmapstream-2018}
Snyk.
\newblock {SNYK-JS-FLATMAPSTREAM-72637: Malicious Package in flatmap-stream},
  November 2018.
\newblock Accessed: 2024-06-09.

\bibitem{sonatype2024}
Sonatype.
\newblock {CVE-2024-3094: The Targeted Backdoor Supply Chain Attack Against xz
  and liblzma}, 2024.
\newblock Accessed: 2024-10-30.

\bibitem{es1}
Ayrton Sparling et~al.
\newblock {Event-Stream, GitHub Issue 116: I don't know what to say.}
\newblock \url{https://github.com/dominictarr/event-stream/issues/116}, 2018.
\newblock Accessed: 2018-12-18.

\bibitem{stokes2019scriptnet}
Jack~W Stokes, Rakshit Agrawal, Geoff McDonald, and Matthew Hausknecht.
\newblock {Scriptnet: Neural static analysis for malicious javascript
  detection}.
\newblock In {\em MILCOM 2019-2019 IEEE Military Communications Conference
  (MILCOM)}, pages 1--8. IEEE, 2019.

\bibitem{tam2024letspeakfreelystudy}
Zhi~Rui Tam, Cheng-Kuang Wu, Yi-Lin Tsai, Chieh-Yen Lin, Hung yi~Lee, and
  Yun-Nung Chen.
\newblock {Let Me Speak Freely? A Study on the Impact of Format Restrictions on
  Performance of Large Language Models}, 2024.

\bibitem{tarr2018eventstream}
Dominic Tarr.
\newblock {event-stream: Commit e3163361fed01384c986b9b4c18feb1fc42b8285}.
\newblock \url{
  https://github.com/dominictarr/event-stream/commit/e3163361fed01384c986b9b4c18feb1fc42b8285
  }, 2018.

\bibitem{harp:ccs:2021}
Nikos Vasilakis, Achilles Benetopoulos, Shivam Handa, Alizee Schoen, Jiasi
  Shen, and Martin~C. Rinard.
\newblock {Supply-Chain Vulnerability Elimination via Active Learning and
  Regeneration}.
\newblock In {\em Proceedings of the 2021 ACM SIGSAC Conference on Computer and
  Communications Security}, CCS '21, New York, NY, USA, 2021. Association for
  Computing Machinery.

\bibitem{lya:fse:2021}
Nikos Vasilakis, Grigoris Ntousakis, Veit Heller, and Martin~C. Rinard.
\newblock {Efficient Module-Level Dynamic Analysis for Dynamic Languages with
  Module Recontextualization}.
\newblock In {\em ACM Joint Meeting on European Software Engineering Conference
  and Symposium on the Foundations of Software Engineering (ESEC/FSE)}, page
  1202–1213, 2021.

\bibitem{mirccs2021}
Nikos Vasilakis, Cristian-Alexandru Staicu, Grigoris Ntousakis, Konstantinos
  Kallas, Ben Karel, Andr\'e DeHon, and Michael Pradel.
\newblock {Preventing Dynamic Library Compromise on \node via RWX-Based
  Privilege Reduction}.
\newblock In {\em ACM Conference on Computer and Communications Security
  (CCS)}, pages 1821--1838, 2021.

\bibitem{software_supply_chain_cost}
Cybersecurity Ventures.
\newblock {Software Supply Chain Attacks to Cost the World \$60B in 2025},
  2023.
\newblock Accessed: 2024-11-05.

\bibitem{wang2018combined}
Yao Wang, Wandong Cai, Pin Lyu, and Wei Shao.
\newblock {A combined static and dynamic analysis approach to detect malicious
  browser extensions}.
\newblock {\em {Security and Communication Networks}}, 2018(1):7087239, 2018.

\bibitem{wei2023chainofthought}
Jason Wei, Xuezhi Wang, Dale Schuurmans, Maarten Bosma, Brian Ichter, Fei Xia,
  Ed~Chi, Quoc Le, and Denny Zhou.
\newblock {Chain-of-Thought Prompting Elicits Reasoning in Large Language
  Models}, 2023.

\bibitem{willard2023efficientguidedgenerationlarge}
Brandon~T. Willard and Rémi Louf.
\newblock {Efficient Guided Generation for Large Language Models}, 2023.

\bibitem{npm_attack_ethereum}
Ryan~Daws Williams.
\newblock {NPM Supply Chain Attack Targets Ethereum Blockchain}, 2023.
\newblock Accessed: 2024-11-05.

\bibitem{wu2023chain}
Dingjun Wu, Jing Zhang, and Xinmei Huang.
\newblock {Chain of thought prompting elicits knowledge augmentation}.
\newblock {\em {arXiv preprint arXiv:2307.01640}}, 2023.

\bibitem{xu2024largelanguagemodelscyber}
Hanxiang Xu, Shenao Wang, Ningke Li, Kailong Wang, Yanjie Zhao, Kai Chen, Ting
  Yu, Yang Liu, and Haoyu Wang.
\newblock {Large Language Models for Cyber Security: A Systematic Literature
  Review}, 2024.

\bibitem{xu2013jstill}
Wei Xu, Fangfang Zhang, and Sencun Zhu.
\newblock {Jstill: mostly static detection of obfuscated malicious javascript
  code}.
\newblock In {\em Proceedings of the third ACM conference on Data and
  application security and privacy}, pages 117--128, 2013.

\bibitem{llmfool:2023}
Xilie Xu, Keyi Kong, Ning Liu, Lizhen Cui, Di~Wang, Jingfeng Zhang, and Mohan
  Kankanhalli.
\newblock {An LLM can Fool Itself: A Prompt-Based Adversarial Attack}.
\newblock Oct 2023.

\bibitem{yaghmazadeh2018automated}
Navid Yaghmazadeh, Xinyu Wang, and Isil Dillig.
\newblock {Automated migration of hierarchical data to relational tables using
  programming-by-example}.
\newblock {\em {Proceedings of the VLDB Endowment}}, 11(5):580--593, 2018.

\bibitem{Yao_2024}
Yifan Yao, Jinhao Duan, Kaidi Xu, Yuanfang Cai, Zhibo Sun, and Yue Zhang.
\newblock {A survey on large language model (LLM) security and privacy: The
  Good, The Bad, and The Ugly}.
\newblock {\em {High-Confidence Computing}}, 4(2):100211, June 2024.

\bibitem{ye2023comprehensivecapabilityanalysisgpt3}
Junjie Ye, Xuanting Chen, Nuo Xu, Can Zu, Zekai Shao, Shichun Liu, Yuhan Cui,
  Zeyang Zhou, Chao Gong, Yang Shen, Jie Zhou, Siming Chen, Tao Gui, Qi~Zhang,
  and Xuanjing Huang.
\newblock {A Comprehensive Capability Analysis of GPT-3 and GPT-3.5 Series
  Models}, 2023.

\bibitem{leftpad}
Serdar Yegulalp.
\newblock {How one yanked JavaScript package wreaked havoc}.
\newblock \url{
  http://www.infoworld.com/article/3047177/javascript/how-one-yanked-javascript-package-wreaked-havoc.html
  }, 2016.

\bibitem{zhang2023does}
Ying Zhang, Wenjia Song, Zhengjie Ji, Danfeng, Yao, and Na~Meng.
\newblock {How well does LLM generate security tests?}, 2023.

\end{thebibliography}
